\documentclass[journal]{IEEEtran}

 \usepackage[font=normal]{caption}

\usepackage{graphicx}
\usepackage{amsmath}

\usepackage{mathtools}
\usepackage{amsmath,amssymb,amsfonts}
\usepackage{algpseudocode, algorithm}
\DeclarePairedDelimiter\ceil{\lceil}{\rceil}
\DeclarePairedDelimiter\floor{\lfloor}{\rfloor}

\usepackage{algorithm}
\usepackage{algpseudocode}

\algnewcommand\algorithmicto{\textbf{to}}

\usepackage{stfloats}

\usepackage{graphicx}
\usepackage{textcomp}
\usepackage{xcolor}
\usepackage{mathtools}
\usepackage{hyperref}
\usepackage{amsmath}
\usepackage{cite}
\usepackage{subcaption}
\usepackage[normalem]{ulem}
\usepackage{soul} 
\usepackage{booktabs}
\usepackage{amssymb}
\usepackage{pifont}
\newcommand{\cmark}{\color{teal} \ding{51}}%
\newcommand{\xmark}{\color{red} \ding{55}}%

\newcommand{\blue}{\color{black}}

\renewcommand\hl[1]{#1}
\newcommand\hleq[1]{#1}

\usepackage{amsmath}

\def\code#1{\texttt{#1}}
\ifCLASSINFOpdf
\else
\fi
\usepackage{url}

\usepackage{amsmath,amsfonts}
\usepackage{algorithm}
\usepackage{array}
\usepackage{textcomp}
\usepackage{stfloats}
\usepackage{url}
\usepackage{verbatim}
\usepackage{cite}

\begin{document}
%
\title{CARTOS: A Charging-Aware Real-Time Operating System for Intermittent Batteryless Devices}

\author{\IEEEauthorblockN{Mohsen Karimi, Yidi Wang, Hyoseung Kim}
\IEEEauthorblockA{University of California, Riverside}
\IEEEauthorblockA{mkari007@ucr.edu, ywang665@ucr.edu, hyoseung@ucr.edu}
}
\author{
    Mohsen~Karimi,~\IEEEmembership{Member,~IEEE,}
    Yidi~Wang,~\IEEEmembership{Member,~IEEE,}
    Youngbin~Kim,~\IEEEmembership{Member,~IEEE,}
    Yoojin~Lim,~\IEEEmembership{Member,~IEEE,}
    Hyoseung~Kim,~\IEEEmembership{Member,~IEEE}
    \thanks{This work is supported in part by grants from IITP (2021-0-00360), NSF (1943265), and USDA/NIFA (2020-51181-32198).}    
    \thanks{M. Karimi and H. Kim are with the Department of Electrical and Computer Engineering, the University of California Riverside, Riverside, CA 92521, USA (email: \href{mailto:mkari007@ucr.edu}{mkari007@ucr.edu};  \href{mailto:hyoseung@ucr.edu}{hyoseung@ucr.edu}). 
    Y. Wang is with the Department of Computer Science and Engineering, Santa Clara University, Santa Clara, CA 95053, USA (email: {\href{mailto:ywang665@ucr.edu}{ywang49@scu.edu}}).  
    Y. Kim and Y. Lim are with the Electronics and Telecommunications Research Institute, Daejeon 34129, South Korea (email: \href{mailto:yb.kim@etri.re.kr}{yb.kim@etri.re.kr}; \href{mailto:yoojin.lim@etri.re.kr}{yoojin.lim@etri.re.kr}).
    }

}
\markboth{IEEE TRANSACTIONS ON EMERGING TOPICS IN COMPUTING,~Vol.~11, No.~4, November~2023}
{Karimi \MakeLowercase{\textit{et al.}}: CARTOS: A Charging-Aware Real-Time Operating System for Intermittent Batteryless Devices}

%


\maketitle
\pagestyle{plain}

\begin{abstract}
This paper presents CARTOS, a charging-aware real-time operating system designed to enhance the functionality of intermittently-powered batteryless devices (IPDs) for various Internet of Things (IoT) applications. While IPDs offer significant advantages such as extended lifespan and operability in extreme environments, they pose unique challenges, including the need to ensure forward progress of program execution amidst variable energy availability and maintaining reliable real-time time behavior during power disruptions. To address these challenges, CARTOS introduces a mixed-preemption scheduling model that classifies tasks into computational and peripheral tasks, and ensures their efficient and timely execution by adopting just-in-time checkpointing for divisible computation tasks and uninterrupted execution for indivisible peripheral tasks. CARTOS also supports processing chains of tasks with precedence constraints and adapts its scheduling in response to environmental changes to offer continuous execution under diverse conditions. CARTOS is implemented with new APIs and components added to FreeRTOS but is designed for portability to other embedded RTOSs. Through real hardware experiments and simulations, CARTOS exhibits superior performance over state-of-the-art methods, demonstrating that it can serve as a practical platform for developing resilient, real-time sensing applications on IPDs.

\end{abstract}

\section{Introduction}
\IEEEPARstart{B}ATTERYLESS systems, often called intermittently-powered devices (IPDs), have the potential to revolutionize a variety of applications in Internet-of-Things (IoT), such as smart healthcare, agriculture, and building monitoring. 
These systems harvest energy from ambient sources such as sunlight, heat, and radio signals, charge a small energy buffer, and execute intermittently whenever energy is available. 
\hl{This approach addresses critical sustainability challenges in IoT deployments by eliminating battery waste and enabling maintenance-free operation for decades, which has driven the growing industry interest in IPD-based solutions. 
Especially, IPDs with supercapacitors as their energy buffers offer significant advantages: much longer lifespans than traditional battery-powered systems, deployability in harsh environments (e.g., extreme temperatures) where batteries may fail, smaller form factors, and reduced manufacturing costs
\!\!\mbox{\cite{denby2020orbital,choi2022compiler,hester2017future}}.}

The aforementioned benefits, however, come with several technical challenges. First, given the variability in energy availability from ambient sources, IPDs must ensure the forward progress of program execution across power failures.
One of the most effective strategies is checkpointing~\cite{Chinchilla2018,Idetic2013,Mementos2011,Mayfly2018}, a method that involves inserting additional code snippets to store the state of a running program in non-volatile memory (NVM). When power is restored, the system can recover from the latest stored state, ensuring continuous progress. However, such static checkpointing introduces a high overhead and can significantly impact the performance of intermittent systems compared to conventional battery-powered embedded systems. Just-In-Time (JIT) checkpointing~\cite{QuickRecall2014,CatNap,Samoyed2019} mitigates this issue by making checkpoints dynamically only when the power goes low. However, it is difficult to be used for tasks with peripheral operations, e.g., sensor access and flash writing, because they cannot be paused and resumed at an arbitrary point, i.e., execution needs to be done atomically.

Secondly, IPDs should provide accurate time behavior across device power cycles that occur due to both intermittent power and limited energy storage.
This is particularly important for periodic sensing applications, which are one of the most in-demand areas of IoT. While many battery-powered devices have been developed using conventional real-time operating systems (RTOS), such as FreeRTOS, $\mu$C/OS, RTEMS, to implement timely operations, they cannot offer forward progress guarantees in IPDs. 
For instance, a power failure during the execution of a sensor-reading or data-logging task can result in incomplete task execution or the loss or corruption of data. Therefore, it is crucial for the RTOS to consider the power consumption of tasks and the energy availability of the system. Furthermore, providing analytical support to understand how these devices will operate under diverse real-world circumstances is of great importance.

In this paper, we propose CARTOS, a charging-aware real-time operating system, as a practical platform to develop energy-resilient and reliable real-time sensing applications on IPDs. To achieve forward progress guarantees without losing efficiency, CARTOS classifies tasks into two types, computation tasks and peripheral tasks, and introduces a mixed-preemption scheduling model. Computation tasks are suspendable and resumable at any point of execution; hence, CARTOS schedules them preemptively and uses JIT checkpointing to store their states. On the other hand, since peripheral tasks are not suspendable or JIT-checkpointable at any arbitrary point, CARTOS checks before their execution if the system has enough energy to complete them, and it schedules them non-preemptively to prevent interruption. Upon this model, CARTOS supports processing chains of tasks, i.e., tasks within an application have precedence constraints, which is common in practical sensing applications.
CARTOS also adapts task scheduling based on environmental changes and ensures correct and timely execution under various conditions. 
Based on our system design, we provide analysis to test the timing behavior of processing chains of computation and peripheral tasks before deployment.  

CARTOS is implemented by introducing new APIs and components to FreeRTOS, which is one of the most widely used embedded RTOS in real-world applications such as Amazon AWS IoT. However, as the design of CARTOS is not specific to FreeRTOS, it is portable to other RTOS as well. From our experiments on real hardware, CARTOS demonstrated several advantages over the state-of-the-art methods.

Section~\ref{sec:background} provides an overview of previous work and establishes essential background information. In Section~\ref{sec:framework}, we introduce the proposed framework. Section~\ref{sec:mixed_preemption_sched_analysis} presents the system model, the proposed mixed-preemption scheduling approach, and offers a detailed scheduling analysis. Section~\ref{sec:Evaluation} presents the results of our evaluation. Lastly, Section~\ref{sec:conclusion} summarizes and concludes this paper.

\section{Background and Related Work}
\label{sec:background}
In this section, we give detailed background on IPD hardware characteristics and discuss state-of-the-art methods developed in the literature. We also review other studies related to energy-harvesting and IPD-based applications.

\subsection{IPD Hardware and Operation Phases}\label{subsec:IPD_OP_Phases}

\begin{figure}
         \centering
         \includegraphics[width=0.48\textwidth]{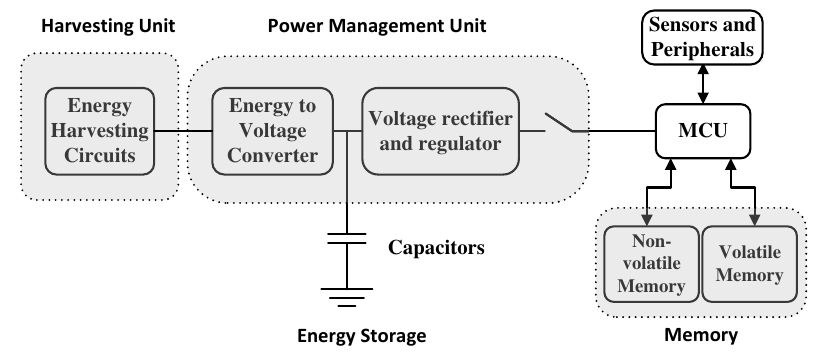}
        \caption{A typical hardware design of IPDs}
        \label{fig:energy_harvesting_BD}
\end{figure}

A typical hardware design of IPDs is illustrated in Fig.~\ref{fig:energy_harvesting_BD}. 
The harvesting unit collects energy from the surrounding environment. It could be a solar cell, RFID receiver, or piezoelectric transducer. 
The power management unit (PMU) is responsible for converting and regulating the energy harvested by the energy harvester into a stable voltage that can be used to power the device's components. It also manages the device's power consumption to ensure that it does not exceed the available energy. 
Energy storage is usually a supercapacitor and is responsible for storing the collected electrical energy so that it can be used later when it is needed. 
The micro-controller unit (MCU) executes the device's software instructions and controls the device's various components.
Memory stores the program code and data that the MCU uses to operate the device. Most IPDs have at least two types of memory: (i) volatile main memory equipped as part of MCU, e.g., SRAM, and (ii) nonvolatile memory to store execution progress during power loss, e.g., FRAM and MRAM. 
IPDs also often include sensors and peripherals, such as temperature, humidity, light sensors, as well as wireless communication modules such as Bluetooth LE and LoRa, to monitor the device's environment, perform certain actions, and report any findings to the user.

\begin{figure}
         \centering
         \includegraphics[width=0.48\textwidth]{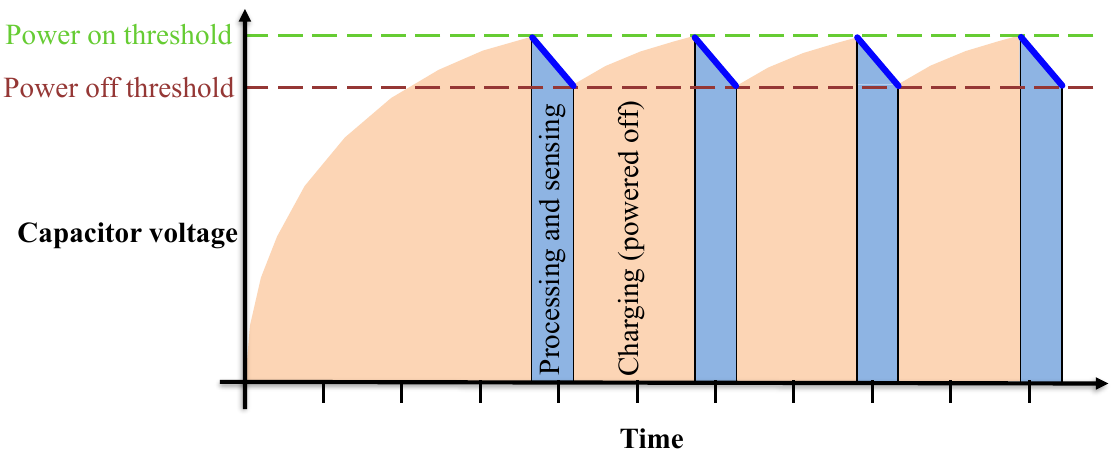}
        \caption{Charging and discharging cycles of an IPD}
        \label{fig:charging_discharging_normal}
\end{figure}

The operation of an IPD can be broken down into the following three phases:
\begin{enumerate}
\item Power-Off Phase: In this initial phase, the device remains entirely powered off. It actively collects energy from the environment using an energy harvester and subsequently stores the gathered energy in a capacitor. 
\item Operation Phase: Once the capacitor attains sufficient voltage, the PMU provides a consistent supply to the device. This enables the MCU to execute tasks. 
The device continues in this phase until it reaches the predefined {\em power-off} threshold (returning to the power-off phase) or {\em low-voltage} threshold (moving to the standby phase discussed next) depending on the design. Many prior IPDs~\cite{colin2016chain,alpaca2017,Chinchilla2018} use only the power-off threshold for simplicity and operate as in Fig.~\ref{fig:charging_discharging_normal}. The MCU starts task execution when the capacitor voltage reaches a specific power-on threshold and continues until the voltage falls below the power-off threshold.
Charging can still occur while the system runs tasks in this phase; however, as the discharging rate typically exceeds the charging rate, the capacitor voltage decreases over time.
\item Standby Phase: In this phase, energy-intensive components like the MCU, sensors, and radio units are either powered down or transitioned into sleep mode. Only the timekeeping ability, e.g., MCU's clock unit or external programmable RTC, remains active \hl{to  maintain} the precise time behavior of tasks~\cite{MohsenIOTJ2021,Celebi2020,Zygarde,karimi2020energy}. The device remains in this state until \hl{the capacitor is sufficiently charged} to return to the operation phase. 
\end{enumerate}

\subsection{Forward Progress Mechanisms}

Various checkpointing mechanisms have been studied to ensure that the progress of a program on IPDs moves forward with data consistency across power failures. Static checkpointing~\cite{Chinchilla2018,Idetic2013,Mementos2011,kortbeek2020time,bhatti2017harvos,kim2023liveness} inserts checkpointing code at the fixed locations, which are determined by either the compiler support or manual programmer efforts. When a checkpoint is reached, the system saves the program's state and relevant data to non-volatile storage, allowing the program to restart from that point in case of a power failure. Furthermore, new programming models have been studied to create checkpoints as part of the program's execution flow~\cite{Mementos2011,Ink2018,Capybara2018,Colin2018,Mayfly2018,colin2016chain,alpaca2017}.\footnote{In the context of IPDs, these approaches are often referred to as {\em task-based} intermittent systems. However, it is important to note that ``tasks'' in these systems differ from our work: they represent short non-preemptible atomic blocks, not-preemptible threads as in conventional RTOS and our work.} In such models, the programmer needs to decompose a program logic into a collection of idempotent, atomic blocks with explicit input and output data.
Then, the input and output data of each atomic block are recorded in NVM, which serves as checkpoints, e.g., data channels between blocks~\cite{colin2016chain, Mayfly2018}. However, this adds a significant burden to programmers to restructure their programs, which is particularly difficult for long-running computations, e.g., algorithm logic. In addition, programmers have to make sure that each atomic block is small enough to run within one charging cycle.

Just-In-Time (JIT) checkpointing~\cite{Samoyed2019,QuickRecall2014,Hibernus2015,CatNap}, on the other hand, makes checkpoints at runtime by capturing the current state of the system, including MCU register values and volatile data. To do so, the system needs to monitor the remaining energy and \hl{initiate JIT checkpointing when the low-voltage threshold is reached.}\footnote{\hl{JIT checkpointing in IPDs \!\!\mbox{\cite{Samoyed2019,QuickRecall2014,Hibernus2015,CatNap}}, including our work, maintains only the most recent checkpoint, not multiple checkpoint versions, due to inherent storage limitations.}} This can bring performance benefits over static checkpointing by reducing the frequency of checkpointing and simplifying the programming of long-running computations~\cite{Hibernus2015,QuickRecall2014}. Also, the overhead of JIT checkpointing is manageably small due to the use of byte-addressable NVM, e.g., FRAM or MRAM; this means the system does not need to checkpoint data stored in NVM. However, since peripheral operations cannot resume after reboot by recovering just the MCU's state, JIT checkpointing cannot be used for them~\cite{Samoyed2019}.

\begin{table*}[t]
\centering
\caption{Comparison of previous work}
\begin{tabular}[t]{lccccccc}
\toprule
\textbf{Methods}&\textbf{JIT checkpointing}&\textbf{Peripherals}&\textbf{Long computations}&\textbf{Real-time scheduling}&\textbf{Energy adaptive}&\textbf{Analysis}\\
\midrule

Mementos\cite{Mementos2011} & \xmark & \xmark & \cmark & \xmark & \cmark& \xmark\\

Chinchilla\cite{Chinchilla2018} & \xmark & \xmark & \cmark & \xmark & \cmark& \xmark\\

Capybara\cite{Capybara2018} & \xmark & \cmark & \xmark & \xmark & \xmark& \xmark\\

Quickrecall\cite{QuickRecall2014} & \cmark & \xmark & \cmark & \xmark & \xmark& \xmark\\

Samoyed\cite{Samoyed2019} & \cmark & \cmark & \cmark & \xmark & \cmark& \xmark\\

InK\cite{Ink2018} & \xmark & \xmark & \xmark & \cmark & \cmark& \xmark\\

CatNap\cite{CatNap} & \cmark & \cmark & \cmark & \xmark & \cmark& \xmark\\

Celebi\cite{Celebi2020}& \xmark & \xmark & \cmark & \cmark & \cmark& \xmark\\

Rtag\cite{MohsenIOTJ2021} & \xmark & \cmark & \xmark & \cmark & \xmark& \cmark\\

Karimi et al.\cite{MohenRTCSA2022} & \xmark & \cmark & \xmark & \cmark & \cmark& \cmark\\\midrule

CARTOS (this work)& \cmark & \cmark & \cmark & \cmark & \cmark& \cmark\\

\bottomrule
\end{tabular}
\label{table:methodComparison}
\end{table*}%

\subsection{IPD Task Scheduling}
Task scheduling for batteryless devices can be categorized into two main types: \textit{real-time} and \textit{reactive} scheduling. 
Real-time scheduling focuses on executing periodic tasks with predictable charging behavior. These approaches aim to meet the deadlines of a specific set of tasks by assuming a certain amount of energy supply\cite{MohsenIOTJ2021, MohenRTCSA2022, Celebi2020}. By doing so, they can guarantee tasks to be scheduled in a predictable manner as long as the assumed energy model holds. On the other hand, reactive scheduling focuses on minimizing the overall response times of event-driven tasks~\cite{CatNap, Ink2018}. It starts task execution when the device's energy reaches the predefined power-on threshold and stops when it drops below the power-off threshold. These approaches are usually more flexible and may perform better with changing environmental conditions than the real-time approaches, but cannot ensure predictability and consistent performance due to their best-effort nature.

\subsection{Related Work}
\label{subsec:related_work}

Table~\ref{table:methodComparison} presents a comprehensive comparison of representative existing methods designed to address task scheduling on IPDs.
As we already characterized prior work by category, we will discuss specific details of some of the work and additional related work below.

Among static checkpointing approaches, Mementos~\cite{Mementos2011} proposes a method to automatically insert checkpointing code to the program at compile time. To reduce the cost of checkpointing, it also proposes to estimate the remaining energy and perform checkpointing only when the energy goes below a certain threshold. 
Chinchilla~\cite{Chinchilla2018} also takes a similar approach. 
Capybara~\cite{Capybara2018} relies on the programming model support for forward progress, and introduces a hardware-software co-design solution that includes a set of capacitors to handle atomic blocks with different energy requirements. 
To address the limitations of programming model-based approaches that require the user to divide program logic into a series of atomic blocks, Immortal Threads~\cite{yildiz2022immortal} takes a compiler-assisted approach that transforms conventional threaded programs by inserting micro checkpoints. This allows the user to write programs in a typical multithreaded manner and enjoy preemption at the boundaries of checkpoints.

Quickrecall~\cite{QuickRecall2014} is a JIT checkpointing system. It detects low-voltage threshold and checkpoints preemptible tasks so that they can be resumed in next power cycle. However, it does not support peripheral tasks and restarts them in next power cycle if the power goes down during their execution. Samoyed~\cite{Samoyed2019} addresses this issue by introducing the {\em undo-logging} mechanism, which basically adds static checkpoints for peripheral operations. However, this approach is applicable only to the peripherals that ``do not have internal non-volatile state and are arbitrarily restartable''~\cite{Samoyed2019} because of the idempotency requirement.

Several approaches have been proposed to support peripheral devices in IPDs. Sytare~\cite{Sytare} allows users to write additional code to explicitly store peripheral states in NVM. RESTOP~\cite{RESTOP} retains peripheral states by recording data transmission between MCU and peripherals, which incurs high overhead. EaseIO~\cite{EaseIO} takes a programming model-based approach that allows users to annotate I/O operations and transforms the program code based on this information. These methods offer various trade-offs between programmer effort, system overhead, and generality of peripheral support.

For real-time scheduling on IPDs, Celebi~\cite{Celebi2020} proposes a scheduler that optimizes task scheduling based on known charging patterns. However, this method is primarily designed for independent computational tasks and may not handle peripheral tasks correctly because the scheduler does not ensure the atomicity and idempotency of task execution, e.g., the scheduler can stop a task at an arbitrary point. Karimi at al.~\cite{MohsenIOTJ2021} introduce a real-time scheduler designed to atomically execute independent periodic tasks with various energy requirements and provide a formal analysis of task schedulability for a fixed charging rate of the system. 
The same authors~\cite{MohenRTCSA2022} present an energy-adaptive scheduler to minimize the age of information. While these approaches contribute to enabling real-time operations on IPDs, none of them are sufficient to handle real-world applications that consist of a chain of peripheral access and long-running computations. 

There are also other previous studies that focus on intermittent neural networks~\cite{BatterylessNN,Zygarde,MII}, multiple energy buffers~\cite{Coulombs}, energy-aware memory mapping~\cite{HybridMem}, communication~\cite{wardega2022opportunistic}, design space exploration~\cite{kim2024rapid}, and zero-power timekeeping~\cite{TimeKeeping}. They contribute to expanding the application scope of IPDs and our work does not conflict with them.

\label{sec:framework}
\begin{figure}[t]
         \centering
         \includegraphics[width=0.48\textwidth]{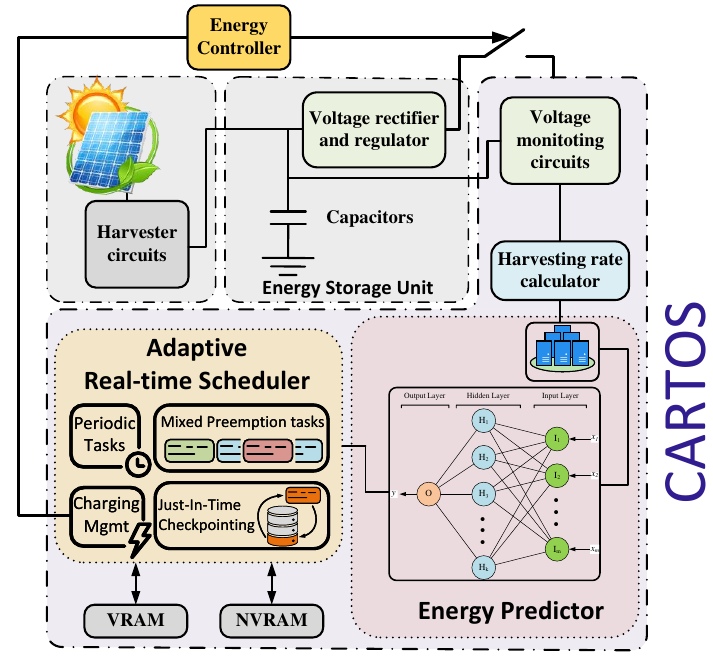}
        \caption{System Framework}
        \label{fig:system_framework}
\end{figure}

\section{CARTOS Framework}

This section presents our proposed CARTOS framework. 
Fig.~\ref{fig:system_framework} depicts the overview of the framework. CARTOS targets typical IPD hardware as discussed in Sec.~\ref{subsec:hardware_setup} and utilizes existing power monitoring and control capabilities of IPDs (right upper corner of the figure). CARTOS includes an adaptive real-time scheduler that is built upon our mixed-preemption scheduling model to support applications with a chain of peripheral and long-running computational tasks. 
CARTOS also includes a lightweight neural network-based energy (charging rate) predictor to adapt to changing environmental conditions. In the rest of this section, we will first describe the hardware setup for CARTOS and then present the major software components of our framework.


\subsection{Hardware Setup}
\label{sec:hardware_setup}
CARTOS targets an IPD that is equipped with an MCU capable of running a conventional RTOS such as FreeRTOS. The IPD is expected to have both volatile memory (SRAM) and non-volatile memory (NVM). Although we used Ambiq Apollo4 Evaluation board \cite{Apollo4_board} which is equipped with an ARM Cortex-M4 MCU in our implementation, there is no other restriction imposed on MCUs by CARTOS. 

While different energy sources can be used, we primarily consider solar energy as a harvesting source. Hence, we assume the IPD has a solar harvesting unit that collects energy from the sun and converts it into usable electrical energy. 
The harvested energy is then stored in a capacitor to drive the IPD. 

To utilize all three operational phases discussed in Sec.~\ref{subsec:IPD_OP_Phases} and support our scheduling behavior, CARTOS assumes that IPD hardware provides the ability to (i) monitor the current charging rate and the remaining energy level of the capacitor, and (ii) cut off power supply to the processing part of the board, e.g., MCU, NVM, sensors, etc., for a requested interval. Item (i) is the voltage monitoring circuits that already exist in typical IPD hardware, and item (ii) is the energy controller shown as the yellow box at the top of Fig.~\ref{fig:system_framework}. Those can be integrated with other hardware components on the same board or can be realized in a separate board, as we did in our implementation (see Sec.~\ref{subsec:hardware_setup}). 

\begin{figure}[t]
	\centering
	\includegraphics[width=0.46\textwidth]{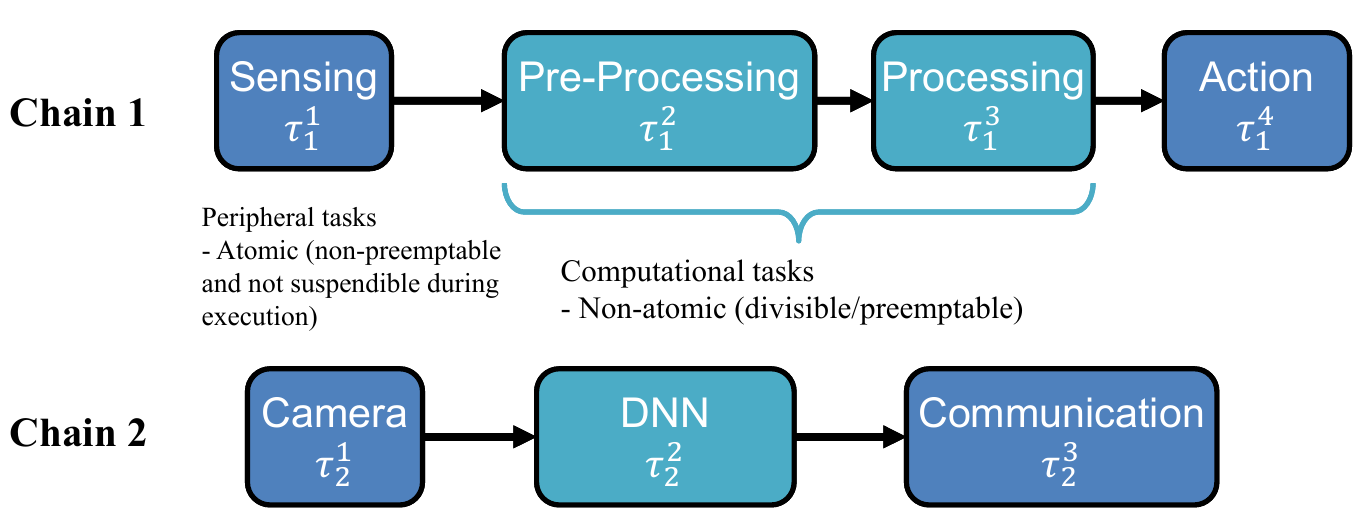}
	\caption{An example of multiple chains with different tasks}
	\label{fig:chains_BD}
\end{figure}

\subsection{Task Chains and Mixed Preemption Model}
CARTOS supports periodic applications that consist of a processing chain of tasks, which is a common design pattern found in sensing and intelligent applications~\cite{hyunjong_chain,sobhani2023timing,colin2016chain,kim2010decentralized,delgado2021optimal}. Fig~\ref{fig:chains_BD} illustrates two processing chains, one comprising four tasks and the other with three tasks. Each task $\tau_i^j$ on a chain~$i$ can be either a peripheral or computation task.\footnote{Detailed chain and task parameters needed for analysis will be provided in Sec.~\ref{sec:mixed_preemption_sched_analysis}.} The chain can start execution when its first task is released (e.g., the periodic sensing task $\tau_1^1$ on chain 1) and the subsequent tasks are eligible to run only when their preceding task finishes (e.g., the action task $\tau_1^4$ becomes ``ready'' after the completion of the processing task $\tau_1^3$). We assume that the first task of a chain is a periodic task, triggered by either a timer or an external event. Hence, all other tasks of the same chain follow the same periodic pattern with precedence constraints among them. 
Chains have deadlines to meet. We consider constrained deadlines (deadline $\le$ period) due to their wide acceptance. A formal model of tasks and chains will follow in Sec.~\ref{sec:mixed_preemption_sched_analysis}.

\subsubsection{Mixed-Preemption Model}
Recall that processing chains can consist of a mixture of peripheral and computation tasks. Peripheral tasks need to be executed {\em atomically}, especially when the corresponding hardware maintains internal state~\cite{Samoyed2019}. 
To ensure the atomic execution of peripheral tasks, CARTOS schedules peripheral tasks non-preemptively, i.e., once selected for scheduling, it will continue with no preemption from other higher-priority tasks. However, this is not enough for atomicity. The adaptive scheduler of CARTOS (Sec.~\ref{subsec:adaptive_scheduler}) waits until enough energy is charged to complete the execution of a peripheral task for this period before scheduling that task. This guarantees that the system does not turn off at least until the completion of the peripheral task, thereby achieving correctness. On the other hand, computational tasks are scheduled preemptively, just like other RTOS, and CARTOS manages their forward progress by JIT checkpointing. To summarize, peripheral and computational tasks in CARTOS have the following characteristics:
\begin{itemize}
    \item Peripheral tasks: atomic, non-preemptible (i.e., not suspendable while it is actively executing; unable to JIT checkpoint)
    \item Computational tasks: non-atomic, preemptible (i.e., suspendable at any time; can be JIT checkpointed)
\end{itemize}

This mixed-preemption model is the basis for our scheduler presented in the next section. The type of a task (atomic or non-atomic) is configurable when the user creates a task. For ease of presentation, we will use non-preemptible, peripheral and atomic tasks interchangeably. The same applies to preemptible, computational, and non-atomic tasks. 

\subsubsection{Memory Model}

Peripheral (atomic) tasks use SRAM for data and stack segments since their memory states need to be initialized upon reboot. In contrast, computational (non-atomic) tasks can utilize NVM for data segments to preserve data across power failures and reduce JIT checkpointing overhead. 
Stack areas of non-atomic tasks could also be allocated in NVM, but our implementation allocates them in SRAM for simplicity and because their memory footprint is relatively small. 
OS-level data structures are maintained in SRAM, with those requiring persistency (e.g., timer events, contexts of non-atomic tasks, and profiling data) being selectively checkpointed. We assume static memory allocation only, as dynamic memory allocation is typically prohibited in real-time systems due to timing unpredictability, and is not considered in our framework.

\begin{figure}[htb]
	\centering
	\includegraphics[width=0.46\textwidth]{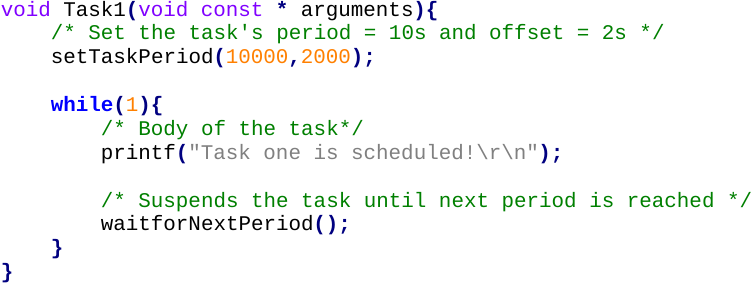}
	\caption{Periodic task example in CARTOS}
	\label{fig:periodic_task_example}
\end{figure}

\subsubsection{Support for Periodic Tasks and Chains}
CARTOS offers APIs to assist the programming and execution of periodic tasks. Users can define tasks as periodic and specify their required periods and release offset based on the application's demands. Specifically, the APIs are:
\begin{itemize}
    \item \code{setTaskPeriod(period, offset)}: This makes the calling task run periodically after the given initial release offset. The second parameter is optional.
    \item \code{waitForNextPeriod()}: This tells the kernel that the calling task has finished its execution for the current period. The task will be suspended immediately and will be released (ready state) at the start of the next period. 
\end{itemize} Fig.~\ref{fig:periodic_task_example} depicts the example of a periodic task in CARTOS. \code{setTaskPeriod()} is called at the beginning, declaring that it is a periodic task. The task has a while-loop where each iteration represents a job for each period. When the task reaches the end of execution in the current period, it calls \code{waitForNextPeriod()}. Then, the scheduler removes it from the ready queue and sets the system timer to insert the task back to the ready queue at the start of the next period. 

For chained task execution, CARTOS also allows the user to create chain objects in the kernel and assign periodic tasks to specific chains. While chain objects are not schedulable entities, they help control the execution order of tasks within a chain. Once tasks are assigned to a chain, these tasks are removed from the ready queue and made inactive. Chain execution commences with the release of the first task, which is triggered by either a timer or an external event as we explained earlier. When the first task finishes its job through the invocation of \code{waitForNextPeriod()}, the subsequent task in the chain is set ready and becomes eligible for execution by the scheduler. This continues until the last task of the chain. Consequently, if all tasks within the chain can complete execution within one chain period (equivalent to the period of the first task), only one task from the chain is ready at any given time. This approach enforces the precedence constraints within a chain while allowing our scheduler to follow the task-level priority-based scheduling policy.


\subsection{Adaptive Real-Time Scheduler}\label{subsec:adaptive_scheduler}

The real-time scheduler of CARTOS is designed to efficiently utilize available energy while supporting both peripheral (atomic) and computational (non-atomic) tasks. 
In doing so, it prioritizes the execution of higher-priority ready tasks over those with lower priority, a crucial feature for meeting timing requirements in RTOS. This priority-based approach is particularly important for systems requiring RTOS in their design, because when not all deadlines can be met due to unexpected workload surges or resource shortages, they need predictability in determining which tasks will miss their deadlines first. By adhering to the priority-based policy, CARTOS minimizes the risk of missing high-priority deadlines, thereby enhancing the system’s overall performance and reliability.

\begin{figure}[t]
	\centering
	\includegraphics[width=0.4\textwidth]{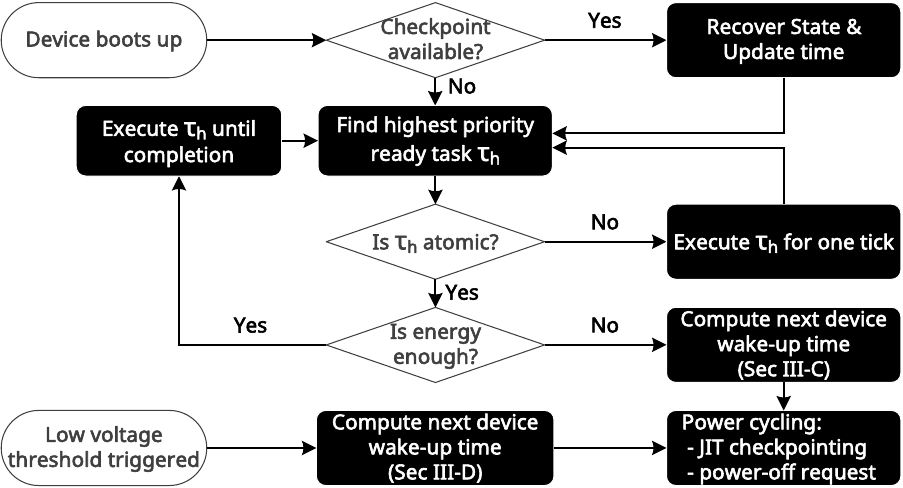}
	\caption{Scheduler operation flowchart}
	\label{fig:scheduler_flowchart}
\end{figure}

Fig~\ref{fig:scheduler_flowchart} illustrates the operation flow of the scheduler from the device boot-up. 
Upon boot-up, the scheduler checks for the presence of any valid JIT checkpoint in NVM. If a checkpoint is found, the device restores tasks from this checkpoint and updates the system time and timer events, including those of \code{waitForNextPeriod()} from periodic tasks.

For scheduling tasks, the scheduler first finds the highest-priority task $\tau_h$ from the ready queue. It is worth noting that the scheduler does not need to consider the execution order of tasks within chains here, because such precedence constraints are controlled by the chain support explained in the previous section.
Then, the scheduler checks if the highest-priority task $\tau_h$ is atomic (peripheral) or not and the scheduling behavior changes according to the task type.

If $\tau_h$ is non-atomic (computational), the scheduler executes $\tau_h$ for one tick of time (scheduling time quantum; 1 ms in our implementation) and repeats the above procedure to find the highest-priority ready task again for preemptiveness. Note that this is the case for tick-based RTOS. In the case of a tickless system, $\tau_h$ can continue to run until it voluntarily relinquishes the CPU or \hl{is preempted by a higher-priority task}. 

If $\tau_h$ is atomic (peripheral), the scheduler checks the required energy for $\tau_h$'s job execution based on the profiled data, and compares it with the current system energy level, measured through the voltage level of the capacitor (details in Sec. {\ref{sec:charging_mgmt}}).
If the system's available energy is higher than the amount required to ensure the uninterrupted execution of the given atomic task, the scheduler dispatches this task and the task will continue until its job is completed, i.e., non-preemptive execution. Otherwise, it proactively enforces a power cycle to harvest more energy.\footnote{Precisely speaking, the device transitions to the standby phase using the energy controller unit discussed in Sec.~\ref{sec:hardware_setup}.} In doing so, it first estimates the {\em required harvesting time}, $\Delta t$, considering the predicted charging rate and the current voltage level,\footnote{It is worth noting that the harvesting time $\Delta t$ can be smaller than the maximum charging demand of the task ($Q_i^j$, explained in Sec~\ref{sec:charging_mgmt}) because the device's current voltage can be higher than the low-voltage threshold.} and then computes the {\em next device wake-up time} from the standby phase by taking into account both the estimated harvesting time and the future release times of other higher-priority periodic tasks. This is done to ensure that higher-priority tasks are not missed during charging. For example, if a high-priority task $\tau_1$ needs to be released at time $t_1$ while the device is being charged for a low-priority task $\tau_2$, the charging process should be stopped so that the high-priority task $\tau_1$ is scheduled. Therefore, the next wake-up time from the standby phase is obtained by taking the minimum between the required harvesting time and the earliest future release time of higher-priority tasks than $\tau_h$.

Subsequently, the device is powered down (after necessary JIT checkpoint operations for preempted \hl{non-atomic} tasks). The atomic task $\tau_h$ as well as other tasks are scheduled when the device is powered on. It should be noted that such a proactive power-cycling sequence happens before reaching the low-voltage threshold. When the device reaches a low-voltage threshold, the device power cycling is managed by the JIT checkpointing service presented next.

\subsection{JIT Checkpointing Service}
In CARTOS, JIT checkpointing is implemented as a system service task. This service monitors if the capacitor voltage level reaches the low-voltage threshold, which can be implemented as an external interrupt or periodic polling of the voltage monitoring circuits. The JIT service task is launched as part of the boot-up sequence and executes with the highest priority to execute as soon as possible when needed.

\hl{JIT checkpointing occurs only once per power cycling (not every scheduling tick), which happens in two cases as depicted in Fig. {\ref{fig:scheduler_flowchart}}: (i) proactive power cycling before atomic task execution (``Is energy enough $\rightarrow$ No'' path in the figure),~and (ii) reactive power cycling by the low-voltage threshold (``Low voltage threshold triggered'' path). In both cases, the next device wake-up time needs to be computed. Since its calculation in case (i) has been discussed already, we now explain that in case (ii). Recall that} 
the low-voltage threshold is triggered only during non-atomic task execution (since atomic tasks execute only with sufficient energy). Let $\tau_i$ denote this interrupted non-atomic task. To compute the next device wake-up time, the system needs $\Delta t$ to harvest energy for $\tau_i$'s remaining portion. However, higher-priority (HP) tasks may be released during charging. To ensure these HP tasks can preempt $\tau_i$ promptly, the system wakes up at the minimum of: (i) the earliest release time of HP tasks, and (ii) the harvesting time $\Delta t$ for $\tau_i$'s completion. Upon wake-up, the scheduler follows Fig. {\ref{fig:scheduler_flowchart}} to handle the highest-priority ready task at the moment.

Once the next device wake-up \hl{time} is computed, JIT checkpointing is performed. It iterates over all non-atomic tasks in the system, and for each non-atomic task, it stores the task's context (task control block), stack, and SRAM data areas (if exist) to NVM so that those can be restored to SRAM upon the next power-on. The JIT service also stores the current system time and timer events in NVM in order to ensure the correct execution of timer-based operations across power cycles. 

When the device powers up, the boot-up procedure restores non-atomic tasks and updates the system time and timer events. The time update can be done in two ways. If there is an external RTC, e.g., as part of the energy controller, it can simply update the system time based on it. Otherwise, it can estimate the current time by adding the latest checkpointed system time and the programmed power-down duration (which equals the computed device wake-up time). We use the latter approach in our implementation.

\hl{If the low-voltage threshold turns out to be insufficient to complete JIT checkpointing before power loss, the system will detect an incomplete or invalid checkpoint upon next reboot. In this case, affected non-atomic tasks simply restart from their beginning, as if they were executing for the first time, and subsequent scheduling operations are performed normally.
}

\subsection{Charging Management}
\label{sec:charging_mgmt}
Charging management is vital to ensure the correct execution of atomic tasks.
To address this, we assign a voltage threshold $V^j_i$ to each atomic task $\tau^j_i$. This threshold represents the minimum voltage that the system requires before starting the execution of task $\tau^j_i$'s job, thereby ensuring that it can complete before reaching the low-voltage threshold used for JIT checkpointing. The calculation for $V^j_i$ is as follows:
	\begin{equation}\label{eq:task_voltage}
		V^j_i = \sqrt{\frac{2Q^j_i W_S+\mathbb{C} V_{min}^2}{\mathbb{C}}} 
	\end{equation}
	\begin{itemize}
		\item $C^j_i$: Worst-case execution time of a job of the task $\tau_i^j$
		\item $Q^j_i$: Maximum charging demand of a task refers to the amount of time required to charge the capacitor at low voltage ($V_{min}$) to perform $C^j_i$ unit of execution.
		\item $W_S$: Energy-harvesting rate of the system
		\item $\mathbb{C}$: Capacitor size in Farad
		\item $V_{min}$: Low-voltage threshold that triggers JIT checkpointing and power cycling
	\end{itemize}
	Eq.~\eqref{eq:task_voltage} is derived from the well-known equation \cite{halliday2013} on the energy stored in capacitor when it receives a fixed power of $W$ from time $t_1$ to $t_2$, given by 
	\begin{equation}\label{eq:capacitor_charging_discharging}
		\Delta E = \frac{1}{2} \mathbb{C}\left(V_2^2 - V_1^2\right) = W \left(t_2 - t_1\right)
	\end{equation}
	To derive Eq.~\eqref{eq:task_voltage} from the above equation, $(t_2-t_1)$ is substituted with $Q^j_i$, $W$ is substituted with $W_s$, $V_2$ is substituted with $V^j_i$, and $V_1$ is substituted with $V_{min}$.
	
	In Eq.~\eqref{eq:task_voltage}, the values of $C^j_i$ and $Q^j_i$ are specific to each task, while $V_{min}$ and $\mathbb{C}$ are the system parameters given by the user at design time. These parameters should be designed so that for all atomic tasks, $V_i^j \le V_{max}$ where $V_{max}$ is the maximum voltage that can be reached by the given energy harvester and the capacitor's nominal voltage. 
    The charging rate of the system, $W_S$, represents a conservative minimum among observations taken at short intervals (e.g., every second, as in \!\!\mbox{\cite{Celebi2020,Zygarde,MohsenIOTJ2021}}). 
    The worst-case execution time $C^j_i$ (obtained a priori) is stored in the TCB and can be updated at runtime to deal with potential underestimation. If there is a change in $C^j_i$ or $W_S$, the value of $Q^j_i$ for that task is also updated at runtime. The calculation for updating $Q^j_i$ is as follows:
	\begin{equation}\label{eq:charging_task_Q}
		Q^j_i = \frac{\left(W^j_{i}-W_S  \right)  \times C^j_i}{W_S}
	\end{equation}
	where $W^j_{i}$ is the discharging rate of $\tau^j_i$.
	
    When scheduling an atomic task $\tau^j_i$, the scheduler compares current voltage level of the system, ${V_{current}}$, with $V^j_i$. If the current voltage level exceeds $V^j_i$, then the scheduler starts executing the task until its completion. Otherwise, the required harvesting time $\Delta t$ is calculated by the following equation and a power cycle is scheduled:    
	\begin{equation}\label{eq:sleeping_time}
		\Delta t = \frac{\mathbb{C}\left( \hleq{( \min{({V^j_i}, V_{max})})^2} - {V_{current}}^2\right)}{2\cdot W_S} 
	\end{equation}
    Here, $V_i^j \le V_{max}$ holds for atomic tasks, as discussed above.
 
    On the other hand, when scheduling non-atomic tasks, the scheduler does not have to consider the current voltage, as depicted in Fig.~\ref{fig:scheduler_flowchart}. Hence, as long as the current voltage is above $V_{min}$, a non-atomic task can begin execution.
    However, if the voltage drops to or below $V_{min}$ during the execution of the non-atomic task, the JIT checkpointing task is activated and computes the required harvesting time $\Delta t$ using Eq.~\eqref{eq:sleeping_time}. {\blue Specifically, $V_i^j$ is computed for the remaining portion of the task and capped by $V_{max}$ \hl{(i.e., $\min{({V^j_i}, V_{max})}$ in Eq.~{\eqref{eq:sleeping_time}})} so that a long-running non-atomic task can execute over multiple power cycles.}
    It is worth noting that our scheduler design improves the average-case runtime performance of non-atomic tasks by reclaiming unused charges from other tasks. However, in the worst case, the maximum total delay experienced by the task is still up to $Q_i^j$ by definition.
    
    The energy harvesting rate, $W_S$, can change over time depending on environmental conditions. As the required harvesting time and the next device wake-up time rely on this, it is crucial to obtain a precise value of $W_S$ for efficient operations because the over-estimation makes the device turn on unnecessarily late and the under-estimation lets the device turn on and off multiple times until it reaches the required charge level. 
    To address this issue, CARTOS employs a machine learning-based energy prediction model proposed in \cite{MohenRTCSA2022}. This model is a simple yet viable approach for MCU-level devices, providing a reliable prediction of solar harvesting rates by leveraging the temporal dependencies of time-series voltage data. In our implementation, we integrated this model into CARTOS and used a 30-minute interval for our prediction window.

Using a single conservative harvesting rate $W_S$ within each prediction interval may lead to pessimism in highly dynamic environments. To reduce this pessimism, one may consider more sophisticated approaches, e.g., \!\mbox{\cite{Zygarde}} captures the probability of $N$ observations of a specific energy harvesting rate, but this is beyond the scope of our paper.


\subsection{Other RTOS Functionalities}
CARTOS allows utilizing standard RTOS features as it is implemented as an extension to FreeRTOS. However, the intermittent nature of batteryless systems introduces unique challenges to some of the key RTOS functionalities. Below we discuss how CARTOS addresses these challenges.

\subsubsection{Semaphores}
Semaphores are crucial for synchronization between real-time tasks. In an intermittent execution environment, managing and recovering lock states across power cycles is needed to ensure lock integrity across power cycles. CARTOS implements the following strategies:
\begin{itemize}
    \item For non-atomic (preemptible) tasks: CARTOS records the lock state and lock-holder information as part of the JIT checkpointing. Upon reboot, the system recovers this information to enable non-atomic lock holder tasks to resume their critical sections with correct lock states.
    \item For atomic (non-preemptible) tasks: With the charging management of CARTOS, atomic tasks are guaranteed to complete before power failure. But for sanity check, we release any locks held by atomic tasks before shutdown so that they can re-enter critical sections after reboot.
\end{itemize}

\subsubsection{Interrupt Handling for External Events}
To suit the intermittent execution model, all interrupt handlers are modeled as atomic tasks for analytical purposes, reflecting their non-suspendable and non-resumable nature. One of the critical interrupts is the timer interrupt, which is handled using a {\em deferred interrupt handling} approach in FreeRTOS. Hence, the interrupt service routine (ISR) of a timer interrupt checks only the hardware timer and the actual timer management for is delegated to a separate timer task. CARTOS models both the timer ISR and the timer task as atomic tasks such that interrupt handling is accounted for in the system's energy management and scheduling decisions.

\subsubsection{Multithreading}
CARTOS fully supports multithreading based on FreeRTOS's task execution model. In FreeRTOS, each independent thread of execution is represented as a task. CARTOS's adaptive real-time scheduler manages these tasks for intermittent conditions while considering both the priority and energy requirements of tasks, as explained earlier.

\section{Mixed-Preemption Processing Chain Analysis}
\label{sec:mixed_preemption_sched_analysis}

This section gives a schedulability analysis of processing chains running in CARTOS. The ability to check if chains will be able to meet their deadlines is important since such information can be used to predict the performance of the system before deployment and provides an opportunity to further optimize hardware and software designs. In the following, we describe the system model and assumptions made for analysis purposes and provide our analysis for chains of tasks.

\subsection{System Model}
\label{sec_sub:system_model}

We consider a system with $n$ processing chains. A chain~$i$ is modeled as $\Gamma_i:(C_i, T_i, D_i, Q_i)$, where $C_i$, $T_i$, $D_i$, and $Q_i$ are the cumulative worst-case execution time, period, deadline, and total charging demand of one instance (job) of the chain, i.e., $C_i = \sum_{j=1}^{m_i} C^j_i$ and $Q_i = \sum_{j=1}^{m_i} \max(Q^j_i, 0)$. The max term is used to handle a negative value for $Q^j_i$, which can occur when the discharging rate of a task is lower than the energy-harvesting rate of the system.
All chains have constrained deadlines, i.e., $\forall i: D_i \le T_i$. Each chain $\Gamma_i$ is comprised of $m_i$ tasks. Each task is either atomic (non-preemtable) or non-atomic (preemptible), following the mixed-preemption model of CARTOS. A task $j$ of a chain $\Gamma_i$ is denoted as $\tau^j_i: (C^j_i, T^j_i, D^j_i, Q^j_i)$, where $\forall j \le m_i ~|~ T^j_i= T_i$,  i.e., all the tasks within the same chain share the same period as their chain and the execution time of the chain is the sum of the execution of all tasks in that chain. Tasks in a processing chain are ordered in sequence. Hence, for a processing chain to complete, all of its tasks should complete their execution in that order, i.e. $\tau^1_i$ should be the first task to be executed and then $\tau^2_i$ and so on. It is important to note that all atomic tasks should satisfy: $V_i^j \le V_{max}$, as explained in Sec. {\ref{sec:charging_mgmt}}. \hl{For non-atomic tasks, we assume that the low-voltage threshold is sufficient to complete JIT checkpointing before power loss. The timing guarantees from our analysis do not hold when these assumptions are violated. Nonetheless, the analysis is useful before deployment as it provides insight into the system's behavior under the assumed conditions.}

Each chain $\Gamma_i$ has a priority of $\pi_i$, e.g., $\pi_h > \pi_i$ means $\Gamma_h$ is a higher-priority chain than $\Gamma_i$. Tasks within a chain share the same task-level priority and they are executed in order. 

We use $R_i$ to denote the worst-case response time of a chain $\Gamma_i$.
A system with $n$ processing chains is said to be {\em schedulable} when the response times of all the chains are smaller than or equal to their deadlines, i.e., $\forall i \le n ~|~ R_i\le D_i$. 

In real-world scenarios, the energy harvesting rate of the system and the charging demand of tasks can vary with the availability of energy resources. However, especially for the solar energy we consider in this work, the changes are relatively slow (e.g., over several tens of minutes and hours) and do not typically happen within a chain period or the hyperperiod of chains. Therefore, a fixed energy harvesting rate is considered for analysis purposes while our runtime system can still adapt to such changes.

\subsection{Schedulability Analysis}
\label{sec_sub:sched_analysis}

Recall our mixed-preemption scheduling model. Atomic tasks are non-preemptively scheduled, so once started, they cannot be preempted by any other tasks. On the other hand, non-atomic tasks are preemptible by higher-priority tasks during their execution. To analyze the worst-case response time of a chain comprised of such tasks, we extend two existing methods: the task-level mixed preemption analysis without energy constraint~\cite{regehr2002scheduling}, and the non-preemptible IPD task scheduling analysis~\cite{MohsenIOTJ2021}, both of which are exact analysis. Our extension takes into account the charging requirements of tasks and the execution order within chains, as follows.

The level-$i$ active period plays an important role in computing the worst-case response time when non-preemptiveness is involved. As such, we first revise the existing calculation of the level-$i$ active period for a task to our processing chain model. 
The level-$i$ active period of a chain $\Gamma_i$ with the charging time demand $Q_i$ can be computed recurrently by:
\begin{equation}\label{eq:level_i_nonpreemptive_withQ}
    L_i^{\left(s\right)} = B_i + \sum_{h:\pi_h\ge \pi_i} \ceil*{\frac{L_i^{\left(s-1\right)}}{T_h}} \left(C_h +Q_h\right)
\end{equation}
where $B_i$ is the blocking time due to a non-preemptible task from a lower-priority chain, $Q_h$ is the cumulative charging demand of a high-priority chain $\Gamma_h$, and $\ceil*{}$ is the ceiling function. $B_i$ can be obtained by
\begin{equation}\label{eq:blocking}
B_i= \max_{\forall l,j \mid\pi_l<\pi_i \land A^j_l=true}\left\{C^j_l \right\}
\end{equation}
where $A_l^j$ indicates the atomicity of a task $\tau_l^j$, i.e., $A_l^j=true$ if $\tau_l^j$ is atomic (non-preemptible) and $A_l^j=false$ if $\tau_l^j$ is non-atomic (preemptible). The iteration of Eq.~\eqref{eq:level_i_nonpreemptive_withQ} starts with $L_i^{\left(0\right)} = B_i + C_i$ and stops when $L_i^{\left(s\right)}=L_i^{\left(s-1\right)}$ (converged) or $L_i^{\left(s\right)}\ge T_H$ where $T_H$ is the hyperperiod of all chains in the system (failed). Here, the charging demand $Q_h$ is added to the execution time $C_h$ because CARTOS can effectively delay
execution by up to this amount when the charge is insufficient.
Recall our scheduler and charging management in Sec.~\ref{sec:framework}. The wake-up time is given as the minimum between the harvesting time $\Delta t$ and the earliest release time, and $Q_i^j$ is greater than or equal to $\Delta t$ by definition (Sec.~\ref{sec:charging_mgmt}). Hence, using $Q_h$ in the analysis provides a safe upper bound for all charging-related delays of individual tasks of $\Gamma_h$ in the worst case.


The use of the level-$i$ active period, $L_i$, allows us to find the worst-case response time of a chain $\Gamma_i$ by considering the response times of all jobs of $\Gamma_i$ within $L_i$. Specifically, 
the chain worst case response time $R_i$ can be obtained by
\begin{equation}\label{eq:response_time_nonpreemptive}
R_i = \max_{k \le K_i} \left\{F_{i,k} - (k-1)T_i\right\}
\end{equation}
where $K_i$ is the number of jobs of $\Gamma_i$ within $L_i$, i.e., $    K_i = \ceil*{\frac{L_i}{T_i}}$, 
and $F_{i,k}$ is the finishing time of the $k^{th}$ job of $\Gamma_i$ in $L_i$. Since tasks within a chain are executed sequentially, $F_{i,k}$ is equal to the finishing time of the last task $\tau_{i,k}^{m_i}$ of $\Gamma_i$, i.e., 
\begin{equation}
F_{i,k} = F_{i,k}^{m_i} 
\end{equation}

If the last task $\tau_{i,k}^{m_i}$ is atomic, due to non-preemptiveness, its latest finishing time $F_{i,k}^{m_i}$ is calculated by
\begin{equation}\label{eq:finish_time_segment_atomic}
    F_{i,k}=S_{i,k}^{m_i} + C_i^{m_i}
\end{equation}
where $S_{i,k}^{m_i}$ is the start time of $\tau_{i,k}^{m_i}$'s execution and we discuss this later.

If the task $\tau_{i,k}^{m_i}$ is non-atomic, $F_{i,k}$ is obtained by the following recurrence, taking into account preemption caused by higher-priority chains during the start and finish time of the task $\tau_{i,k}^{m_i}$:
\begin{multline}\label{eq:finish_time_segment_nonatomic}
    {F_{i,k}}^{\left(s\right)} = S_{i,k}^{m_i} + C^{m_i}_i \\
    + \sum_{h:\pi_h> \pi_i} \left( \ceil*{\frac{{F_{i,k}}^{\left(s-1\right)}}{T_h}} - \left(\floor*{\frac{S_{i,k}^{m_i}}{T_h}}+1\right)\right)(C_h + Q_h)
\end{multline}
where $\floor*{}$ is the floor function. 
The iteration in Eq.~\eqref{eq:finish_time_segment_nonatomic} can start with ${F_{i,k}}^{(0)} = S_{i,k}^{m_i} + C^{m_i}_i$. 

To find the start time of task $\tau_{i,k}^{m_i}$, $S_{i,k}^{m_i}$, we should consider the following factors: (i) blocking caused by lower priority chains, (ii) preemption caused by higher priority chains, (iii) execution time of previous jobs of the chain $\Gamma_i$, (iv) executions of all other tasks in the chain $\Gamma_i$ that should be serviced before the task $\tau_{i,k}^{m_i}$, and (v) all the \hl{extra time} caused by charging. Note that the job of the chain $\Gamma_i$ can be blocked only once during its entire job execution due to the priority-based scheduling of CARTOS. Therefore, for the task $\tau_{i,k}^{m_i}$, the start time $S_{i,k}^{m_i}$ can be computed by the following recurrence:
\begin{multline}\label{eq:starting_time_with_q}
{S_{i,k}^{m_i}}^{(s)}
    = B_i + \left(k-1\right)C_i + \sum_{p=1}^{m_i-1} C_i^p\\
+ \sum_{h:\pi_h>\pi_i}\left(\floor*{\frac{{S^{m_i}_{i,k}}^{(s-1)}}{T_h}} + 1\right)C_h + {\nu_{i,k}}^{(s)}
\end{multline}
where the starting condition is 
\begin{equation}\label{eq:starting_time_s0}
{S_{i,k}^{m_i}}^{(0)} = (k-1)\cdot T_i + B_i + \sum_{p=1}^{m_i-1} C_i^p
\end{equation}
In Eq.~\eqref{eq:starting_time_with_q}, ${\nu_{i,k}}^{(s)}$ is the total amount of time required for $k$ jobs of the chain $\Gamma_{i}$ to acquire their own portions of charge. This can be calculated at each iteration by considering the charging demand of higher-priority chain jobs:
\begin{multline}\label{eq:nu}
{\nu_{i,k}}^{(s)} = k\times Q_i
+\sum_{h:\pi_h>\pi_i}\left(\floor*{\frac{{S^{m_i}_{i,k}}^{(s-1)}}{T_h}} + 1\right)Q_h
\end{multline}


From the above, we can compute the worst-case response time of each chain and test the schedulability of the system.



\section{Evaluation}
\label{sec:Evaluation}
We demonstrate the effectiveness of CARTOS through two sets of evaluations. First, we use our implementation on a real system to compare CARTOS against various state-of-the-art approaches. Second, we conduct schedulability experiments to evaluate our scheduling framework's performance under various conditions, against the best-performing existing approach.

\begin{figure}
         \centering
         \includegraphics[width=0.47\textwidth]{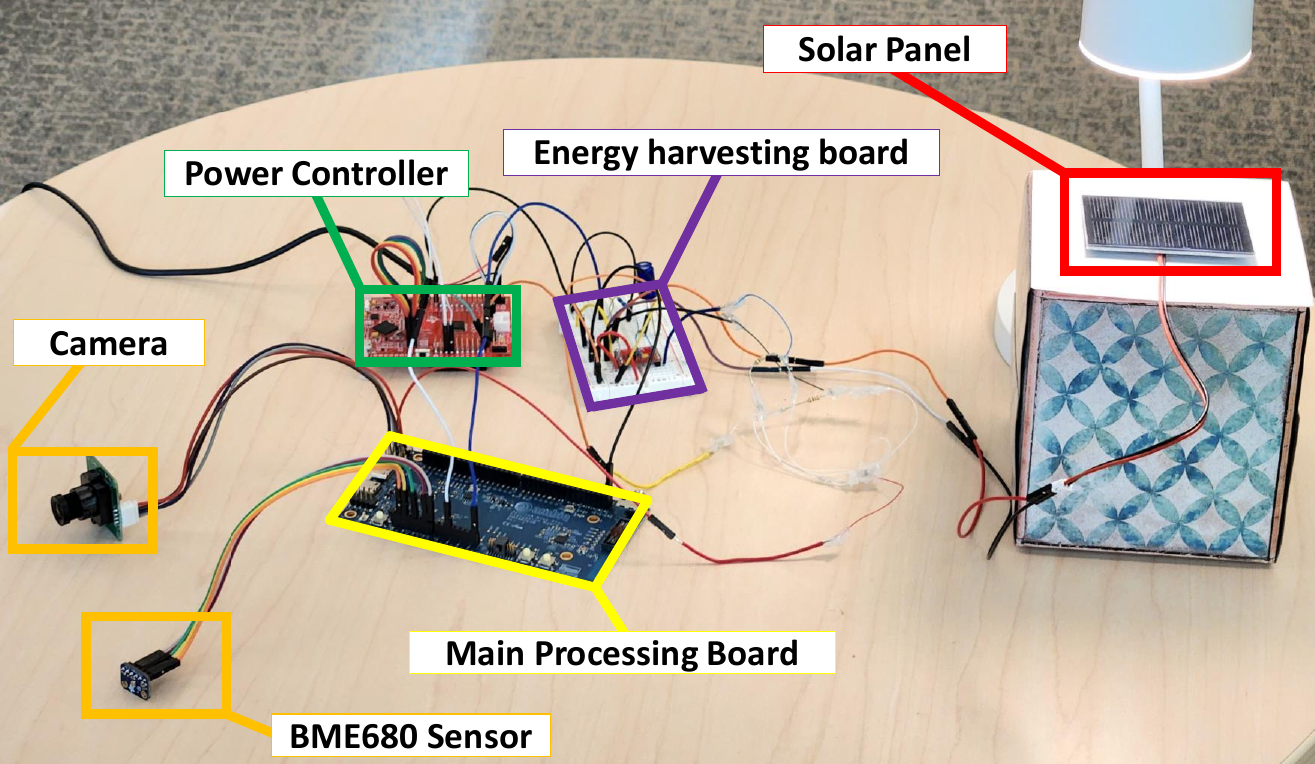}
        \caption{Hardware setup for evaluation}
        \label{fig:experimental_setup}
\end{figure}
\subsection{Hardware Setup}\label{subsec:hardware_setup}
Fig.~\ref{fig:experimental_setup} depicts the hardware setup for our experiments. The main processing board is an Ambiq Apollo4 Evaluation board \cite{Apollo4_board} equipped with a Cortex-M4 Processor, 2MB internal non-volatile MRAM, and 2.75MB low power SRAM.
Power is supplied by an energy harvesting board comprising a 30mF, 100mF, or 470mF capacitor (will be explained in Sec~\ref{subsec:experiments}), LTC3588 energy harvester, and a 2W solar panel. An energy controller board with an ultra-low-power MSP430 MCU is used to control the power supply to the processing board and record experimental results. We have also integrated and configured a camera within the system to facilitate applications related to image processing as well as an environmental sensor to measure temperature, humidity, total volatile compound (TVOC) for environmental sensing applications. In our setup, the power-on and power-off thresholds are 4.04V and 2.9V respectively,  determined by the datasheet when generating 1.8V. 
\hl{The low-voltage threshold is determined empirically as follows: (i) measure the time required for JIT checkpointing (2.57ms in our system, as shown in Table {\ref{table:overhead}}) using the MCU's internal timer (or an oscilloscope with GPIO signals); (ii) measure the voltage drop rate under maximum system load; (iii) compute the required voltage delta $\delta$ as the checkpointing time multiplied by the voltage drop rate; and (iv) obtain the low-voltage threshold as the sum of the power-off threshold, $\delta$, and safety margin. In our implementation, this yielded 3.0V; the 0.1V difference from the power-off provided sufficient energy buffer for JIT checkpointing throughout our experiments.}
The main processing board measures the capacitor voltage at runtime using the MCU's \hl{12-bit internal ADC, which provides a resolution of $\approx$ 0.0015V per step given our voltage range and takes about 1$\mu$s per sample for conversion.}


To control the power connection between the energy harvesting and processing boards, MOSFET transistors are used as a switch. This switch is placed between the regulated power from the energy harvesting board and the processing board and is controlled by the energy controller. The main processing board uses serial communication to request power cut-off from the energy controller for a specific amount of time. The energy controller then sets a GPIO pin low for the specified time and reverts it back to high when the time expires. 
The GPIO pin is connected to the MOSFET transistor that switches the power between the energy harvester and the main processing board. 

While we used an external MCU (MSP430) in the energy controller to facilitate power supply control and experimental data recording, it could be replaced by an ultra-low-power programmable RTC (e.g., 14nA with Ambiq AM0815) or similar device capable of setting a GPIO pin for a programmable duration. In fact, even with MSP430, the power consumption of the energy controller is rather negligible ($<$0.5mW in peak usage) compared to that of the main processing board (9-93mW as shown in Table~\ref{table:Apollo_Exp_task_parameters}). Hence, we did not consider the power overhead of the energy controller. 

The source code of our implementation is publicly available at \url{https://github.com/rtenlab/CARTOS}.

\begin{table}[t]
\centering
\caption{Task parameters for the hardware implementation}
\label{table:Apollo_Exp_task_parameters}
\begin{tabular}{l c c c c c }
\toprule
Task $i$            & $C_i$ (ms)    & $T_i$ (s)     & $W_i$ (mW)    & $\pi_i$   & Preemptible\\ 
\midrule
1: CRC             & 76            & 5             & 9.49          & 7         & Y\\ 
2: Sensor          & 301           & 6             & 57.54         & 6         & N\\ 
3: SHA             & 416           & 8             & 9.8           & 5         & Y\\ 
4: FFT             & 1680          & 10            & 10.02         & 4         & Y\\ 
5: String search   & 3235          & 15            & 10.13         & 3         & Y\\ 
6: Camera          & 3997          & 60            & 93.88         & 2         & N\\ 
7: Basic Math      & 12870         & 120           & 9.59          & 1         & Y\\ 
\bottomrule
\end{tabular}%
\end{table}

\begin{figure*}[t]
     \centering
     \begin{subfigure}{0.329\textwidth}
         \centering
        \includegraphics[width=\textwidth]{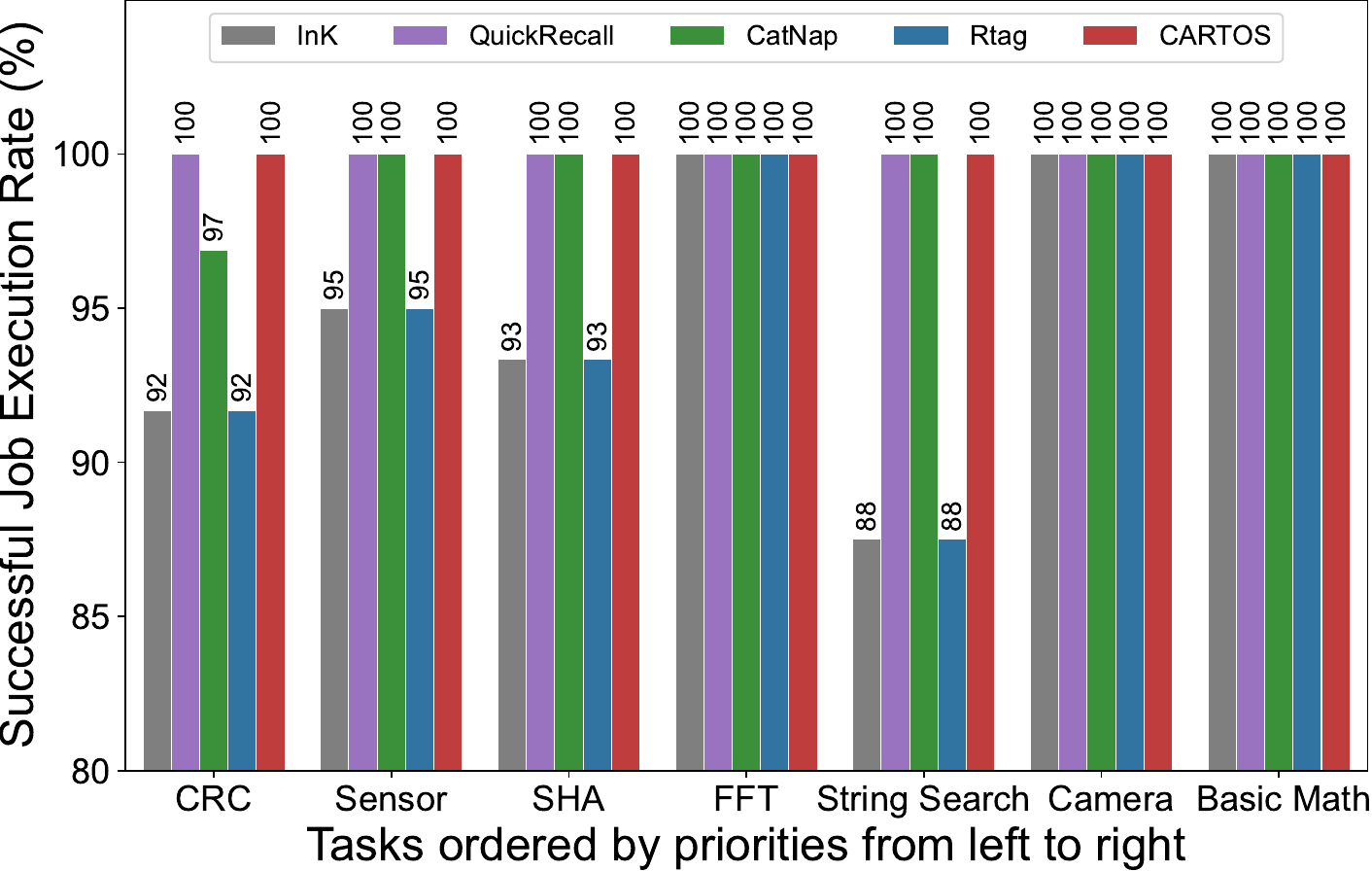}
       \caption{\scriptsize \centering $\mathbb{C}=100mF$, Charging Rate=$\infty$ (Ideal)}
        \label{fig:Exp1_Apollo_Inf_100mF}
    \end{subfigure}
     \begin{subfigure}{0.329\textwidth}
        \centering
        \includegraphics[width=\textwidth]{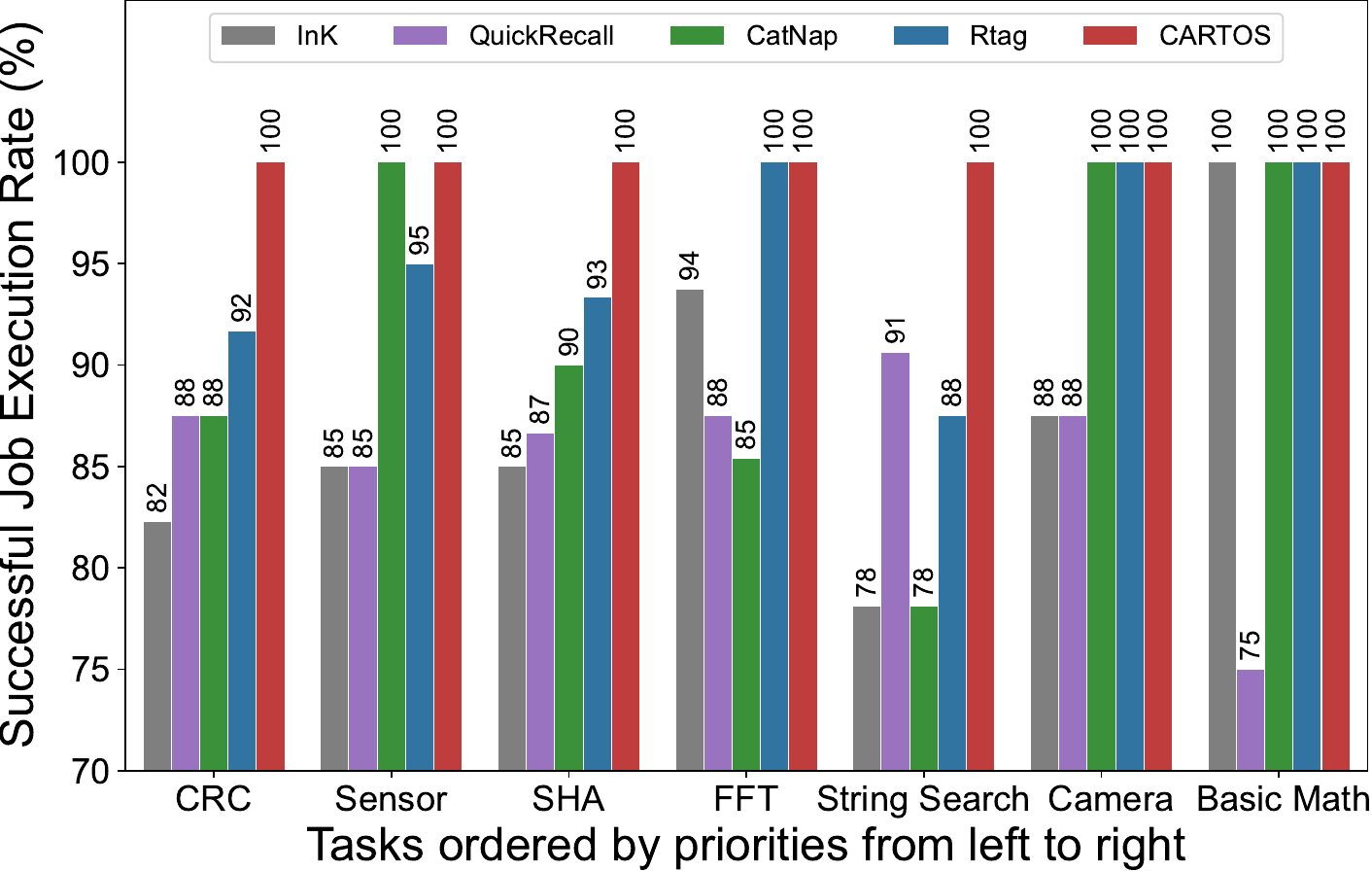}
        \caption{\scriptsize \centering $\mathbb{C}=100mF$, Charging Rate=$15mW$ (Moderate)}
        \label{fig:Exp1_Apollo_15mW_100mF}
    \end{subfigure}
    \begin{subfigure}{0.329\textwidth}
         \centering
        \includegraphics[width=\textwidth]{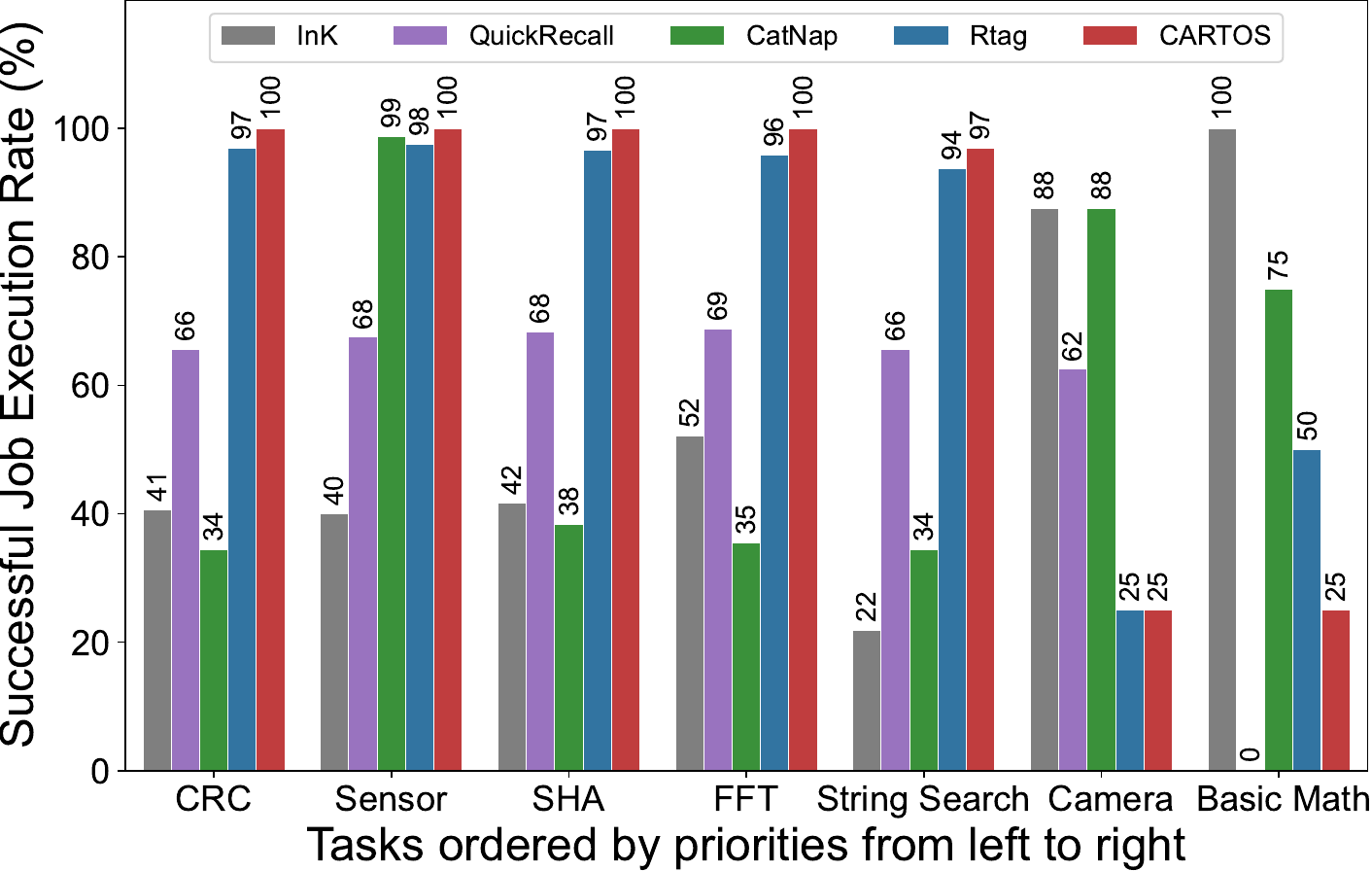}
        \caption{\scriptsize \centering $\mathbb{C}=100mF$, Charging Rate=$8mW$ (Scarce)}
        \label{fig:Exp1_Apollo_8mW_100mF}
    \end{subfigure}    
    \caption{Task execution behavior of the proposed method compared to the other methods in different charging rates}
    \label{fig:Exp1_Apollo}
\end{figure*}

\subsection{Real System Experiments} 
\label{subsec:experiments}

In the following, we evaluate CARTOS against state-of-the-art IPD scheduling methods under various experimental scenarios. All experiments in this section consider only single-task chains since most previous work in this domain does not thoroughly support task chains.\footnote{\blue Recall that chains in our work refer to the sequence of long-running threads, not that of short atomic blocks considered in some prior work~\cite{colin2016chain}.} 
To maintain consistency and repeatability, we employ several distinct harvesting conditions throughout the experiments (explained later). 
Each experiment begins when the capacitor reaches the ``power-on'' threshold and runs for four hyperperiods (480 seconds) to minimize the impact of initial capacitor charge. We used a solar panel and an LED lamp placed at a fixed distance apart. Our measurements show that the solar panel generates 8mW or 15mW of power when the LED lamp is on \textit{low} or \textit{high} setting, respectively.

Table \ref{table:Apollo_Exp_task_parameters} lists the taskset used in our experiments. This taskset includes preemptible tasks from the MiBench benchmark suite \cite{MiBench_Benchmark}, along with two non-preemptible tasks: Sensor reading and Camera capture. Following common practice in real-time embedded systems, we used the rate-monotonic (RM) priority assignment, which assigns higher priorities ($\pi_i$) to tasks with shorter periods. For each task, the worst-case execution time ($C_i$) and average power consumption ($W_i$) were measured offline by executing each task 1000 times to obtain robust measurements. We compare CARTOS against InK \cite{Ink2018}, Quickrecall \cite{QuickRecall2014}, CatNap \cite{CatNap}, and Rtag \cite{MohsenIOTJ2021}. These methods collectively represent the techniques listed in Table~\ref{table:methodComparison} or have been shown to outperform the others. Hence, this provides a comprehensive comparison of the efficacy of CARTOS across various scheduling scenarios.

\subsubsection{Scheduler performance comparison in different charging conditions}
Since IPDs can experience varying charging conditions due to environmental changes, in this experiment, we evaluate our proposed method against related methods under different charging environments. We consider the following three energy harvesting modes:
\begin{itemize}
    \item Ideal: In this mode, the charging rate significantly exceeds the discharging rate of all tasks, causing the capacitor’s voltage to remain at the maximum output of the solar panel (5.8V for the panel used), similar to when the device is plugged into wall power. This situation can occur when the solar panel provides a high amount of energy, either due to its large size and excellent sun exposure or because the tasks consume relatively little power compared to the solar energy available. To achieve this, we directly connected the capacitor terminals to the external power supply of 5.8V.
    \item Moderate: In this mode, the charging rate is high enough to schedule the taskset with an ideal scheduler, i.e., the energy provided by the energy harvester during a hyperperiod is equal to or higher than the sum of the energy consumed by all jobs of all tasks in one hyperperiod. For the given taskset, the charging rate of 15mW is enough to ensure schedulability based on the analysis provided in \ref{sec_sub:sched_analysis}. In this scenario, the utilization of the taskset when considering charging and execution of tasks is calculated to be 97\% (i.e., $\sum (Q_i+C_i)/T_i \approx 0.97$).
    \item Scarce:  In this mode, the charging rate is significantly lower than the requirements of the taskset, making the taskset unschedulable. However, it is important to observe the system's behavior under different methods in this mode, as facing such conditions is unavoidable in IPDs. 
\end{itemize}

InK and Rtag only consider non-preemptive scheduling \cite{Ink2018, MohsenIOTJ2021}. InK and Quickrecall use a best-effort approach \cite{Ink2018, QuickRecall2014}, i.e., execute tasks as soon as the device's voltage reaches the power-on threshold and continue until the voltage drops to the power-off threshold. Therefore, we considered the capacitor size so that in all of the methods, at least one job of each task can complete its execution in one power cycle, i.e., $\frac{1}{2}*\mathbb{C}\times (V^2_{on} - V^2_{off})> \max_i \{(W_i-W_S) \times C_i\}$, where $\mathbb{C}$ represents the capacitance, $V_{on}$ and $V_{off}$ are the power-on and power-off voltage thresholds, $W_i$ is the average power consumption of task $i$, $W_S$ is the power harvesting rate of the system, and $C_i$ is the worst-case execution time of task $i$. With this, the performance metric we considered is the ratio of successfully executed jobs to total jobs released since uncompleted jobs cannot produce correct output in IPDs.  

The results from the experiment are presented in Fig.~\ref{fig:Exp1_Apollo}. Fig.~\ref{fig:Exp1_Apollo_Inf_100mF} shows that, while QuickRecall and CARTOS are able to schedule the taskset without any missed deadlines in the ideal harvesting mode, InK, Rtag, and CatNap fail to do so. For Ink and Rtag, this failure is primarily due to their non-preemptive scheduling approach, which leads to extended and unnecessary blocking times for higher-priority tasks, causing some higher-priority jobs to miss their deadlines. In case CatNap, a small number of failures for the highest-priority task (CRC) is due to its hierarchical scheduling structure, which we analyze further in the next paragraph.

Fig.~\ref{fig:Exp1_Apollo_15mW_100mF} demonstrates the scheduling behavior in moderate harvesting mode. In this scenario, only CARTOS successfully schedules all jobs, while the other methods result in missed deadlines, even for higher-priority tasks.
InK lacks JIT checkpointing support and executes all tasks atomically. 
Therefore, when the voltage drops below the \textit{turn-off} threshold, any running task under InK loses progress and must restart in the next power cycle, thereby wasting energy.
Quickrecall uses a ``best-effort'' approach to task execution, only stopping job execution and performing JIT checkpointing at the \textit{low-voltage} threshold. As a result, it fails to complete some non-preemptible jobs, forcing them to restart in the next power cycle and causing energy waste.
CatNap supports JIT checkpointing for computational tasks and monitors charging requirements for peripheral tasks (`events' in \cite{CatNap}), allowing them to complete atomically without power interruption. However, it takes a hierarchical scheduling approach that strictly prioritizes peripheral tasks over computational tasks, nullifying the actual priority assignment and causing energy starvation for high-priority tasks.
Rtag restricts all tasks to execute atomically, causing long blocking times for higher-priority tasks. It also faces an out-of-order energy consumption issue similar to CatNap, where lower-priority tasks consume energy that could otherwise be used by higher-priority tasks.

CARTOS, on the other hand, benefits from higher charging rates to schedule all jobs without missing any deadlines. Even with lower charging rates, it respects task priorities and provides predictable and robust performance, which is a crucial property sought in RTOS.

Fig.~\ref{fig:Exp1_Apollo_8mW_100mF} shows the scheduling behavior of all methods in the energy scarce condition. In this case, the average harvesting rate of $8mW$ leads to a taskset utilization of 183\% and making it unavoidable for some jobs to miss their deadlines. As shown in the figure, CARTOS is able to successfully finish higher-priority jobs at the expense of lower-priority jobs. However, all other methods fail to complete some of the highest-priority jobs. This shows that CARTOS can effectively schedule real-time tasks in priority order, even when energy is scarce.

\begin{figure*}[ht]
     \centering
     \begin{subfigure}{0.329\textwidth}
        \includegraphics[width=\textwidth]{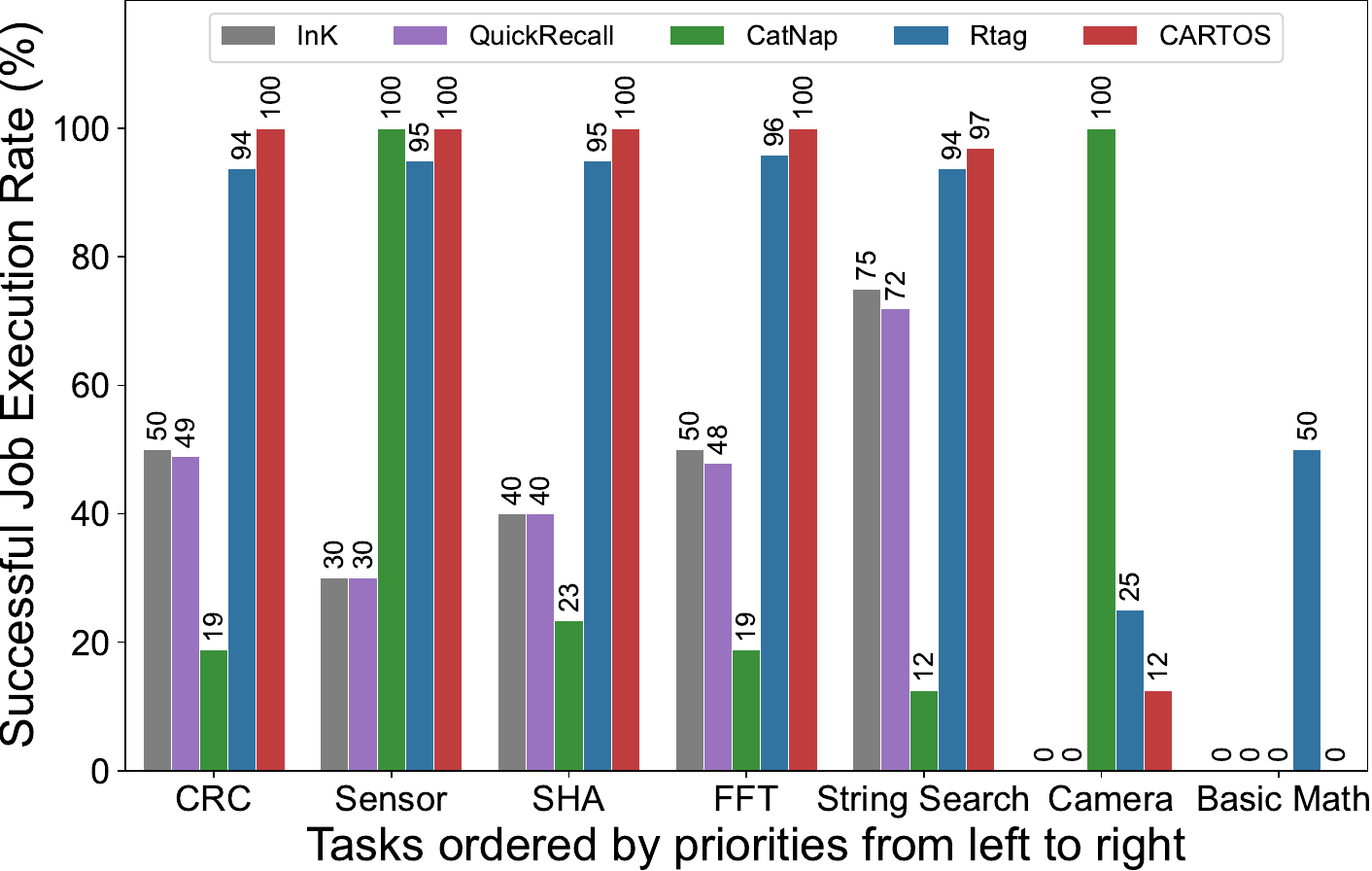}
        \caption{\scriptsize \centering $\mathbb{C}=30mF$, Charging Rate = $8mW$}
        \label{fig:Exp2_Apollo_8mW_30mF}
    \end{subfigure}
     \begin{subfigure}{0.329\textwidth}
        \includegraphics[width=\textwidth]{Figures/Exp_Apollo_8mW_100mF.pdf}
        \caption{\scriptsize \centering $\mathbb{C}=100mF$, Charging Rate = $8mW$}
        \label{fig:Exp2_Apollo_8mW_100mF}
    \end{subfigure}
    \begin{subfigure}{0.329\textwidth}
        \includegraphics[width=\textwidth]{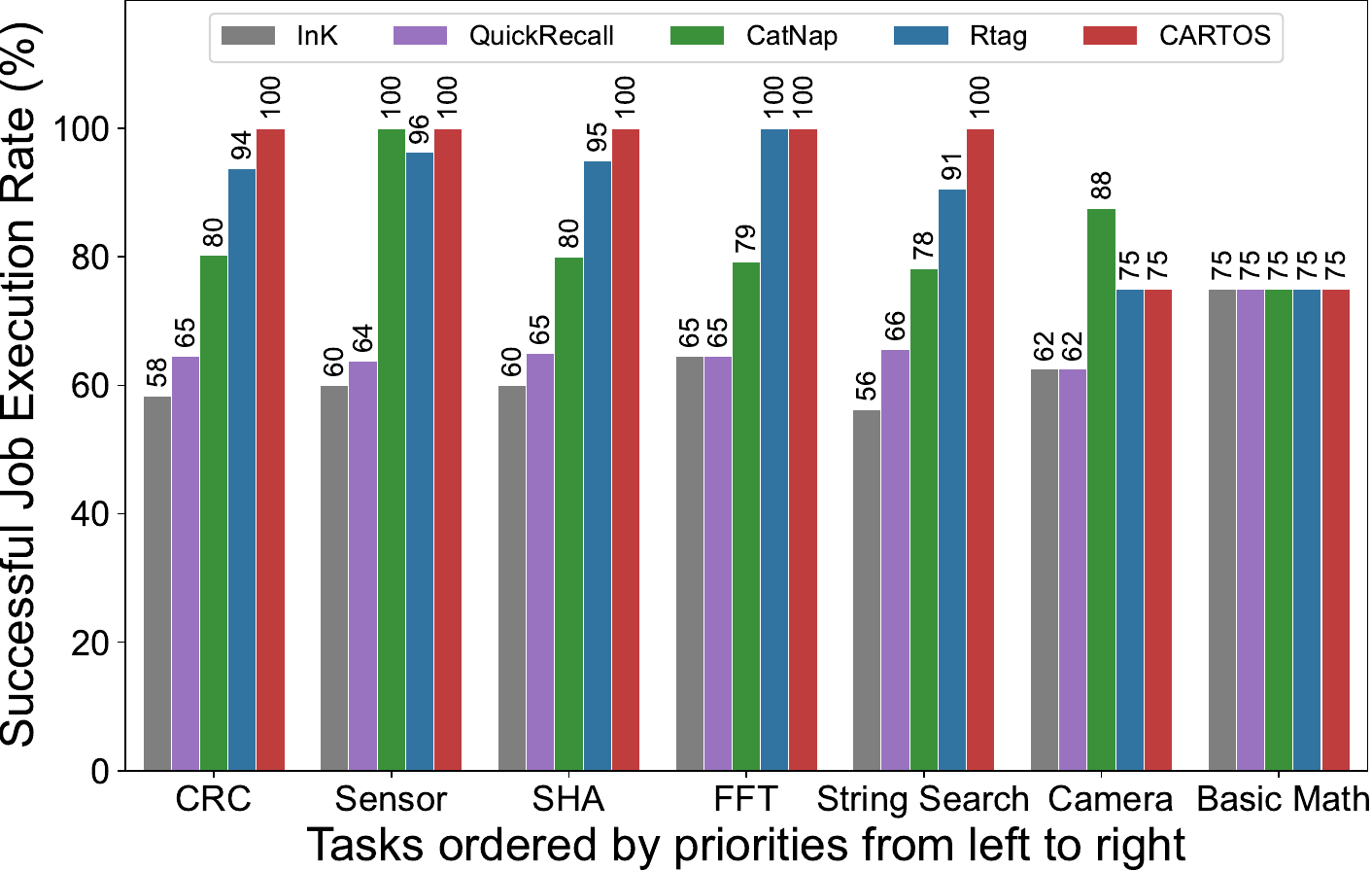}
        \caption{\scriptsize \centering $\mathbb{C}=470mF$, Charging Rate = $8mW$}
        \label{fig:Exp2_Apollo_8mW_470mF}
    \end{subfigure}    
    \caption{\blue Task execution behavior of the proposed method compared to the other methods in different capacitor sizes}
    \label{fig:Exp2_Apollo}

\end{figure*}

\subsubsection{Scheduler performance comparison in different capacitor sizes}
Since most IPDs are designed to be small and cost-efficient, the size of the capacitor plays a significant role. In addition, some methods may perform better with a larger or smaller energy storage capacitor. To address this, we designed a new experiment to evaluate the performance of each method with different capacitor sizes. Similar to the previous experiment, capacitors are charged to the turn-on threshold before starting the schedulers to ensure a fair comparison.
In the first experiment, we determine the smallest capacitor that can be used given the discharging rate of the non-preemptible tasks and the maximum voltage of the solar panel. This ensures that any atomic task can complete at least one job execution before reaching the turn-off voltage if it starts executing the job at maximum voltage. Specifically, we calculate the minimum capacitor size using:
\begin{equation}\label{eq:min_capacitor}
\mathbb{C}_{min}= \frac{\max_{\forall i \mid A_i=true}\left\{C_i \times W_i \right\}}{\frac{1}{2} \times \left( V^2_{max} - V^2_{min}\right)}
\end{equation}
where $C_i$, $W_i$, and $A_i$ are the worst-case execution time, average power consumption, and atomicity of task $i$, respectively. $\mathbb{C}_{min}$ is the minimum capacitor size, $V_{max}$ is the maximum voltage of the solar panel, and $V_{min}$ is the low-voltage threshold.

Considering that our solar panel can generate up to 5.8V and the low-voltage threshold is 3V, the capacitor size is calculated as $\frac{3.997\times 0.09388}{0.5\times (5.8^2-3^2)} \approx 0.030F$. Therefore, we consider $30mF$ as the minimum capacitor size that can serve the current taskset if enough charging is provided. We also set the charging rate to $8mW$ to observe the behavior of each method when the charging rate is lower than expected but the energy storage is large enough. It should be noted that if we choose an optimal charging rate (i.e. $15mW$), having a very large capacitor will produce similar behavior as the ideal mode shown in Fig.~\ref{fig:Exp1_Apollo_Inf_100mF}. This is due to the fact that a large capacitor creates a very large energy buffer that provides energy more than one hyperperiod of task execution. If the charging rate of the system is chosen to be larger than the discharging rate of the taskset in one hyperperiod, the device will never reach the \textit{low-voltage} or \textit{turn-off} threshold, resulting in behavior similar to that observed in Fig.~\ref{fig:Exp1_Apollo_Inf_100mF}.

Fig.~\ref{fig:Exp2_Apollo} shows the experimental results for three capacitor sizes of $30mF$, $100mF$, and $470mF$. In the experiment, even with the smallest capacitor, CARTOS successfully schedules all higher-priority tasks, with only lower-priority tasks missing some jobs. In contrast, all other methods fail to meet deadlines for some of the higher-priority tasks.
Interestingly, some methods perform worse for some tasks with the largest capacitor ($470mF$). This counter-intuitive behavior can be explained by the trade-off between initial energy and charging time.
As the capacitor size increases, the initial energy of the capacitor at $V_{on}$ voltage increases. However, it also increases the time required for the voltage rise from $V_{off}$ to $V_{on}$. 

CatNap and Rtag, which sometimes consume critical charging for lower priority tasks, could schedule tasks better with a larger capacitor size only in the first two periods (half of the experiment). However, they struggle in the latter half when the capacitor approaches $V_{off}$ and they miss more deadlines. InK and Quickrecall, which use the best-effort approach, are severely affected by extremely large capacitors, as the device turns off at $V_{off}$ and takes much longer to reach $V_{on}$, causing very long blackout time and a high number of missed jobs, particularly for higher-priority tasks with smaller periods. On the other hand, CARTOS demonstrates robust performance and predictability in deadline misses. It effectively leverages the larger capacitor and initial charge to execute lower-priority tasks while consistently meeting deadlines for higher-priority tasks even when the capacitor size is small.

As demonstrated in the previous experiments, CARTOS consistently outperforms all other methods across various scenarios. Regardless of the charging rate and capacitor size, CARTOS successfully schedules higher-priority tasks and produces predictable outputs.

\subsubsection{Scheduler and JIT Checkpointing Overhead of CARTOS}
\label{subsec:overhead}
During the experiment, we  measured the overhead caused by JIT checkpointing mechanism for preemptible tasks. The JIT checkpointing overhead was measured to be less than 0.088\% in worst case scenario. It should be noted that the maximum JIT checkpointing overhead occurs when the smallest capacitor and the lowest charging rate were used, causing the largest number of power cycles. For the taskset used in previous experiments, the scheduler overhead was very negligible and could not be precisely measured by our equipment. \hl{Instead, by comparing device uptime and task execution time, we could estimate the scheduler overhead (including voltage monitoring costs) to be less than 1ms ($<$ 0.001\%).}
This shows that CARTOS delivers high-performance task scheduling on IPDs with minimal impact on system resources. We also measured the power consumption overhead associated with writing to and reading from MRAM compared to SRAM. To do so, we conducted 64MB of read/write operation and measured the average power consumption.
Table~\ref{table:overhead} provides the overhead measurements in details.
\begin{table}[H]
\centering
\caption{\blue Breakdown of CARTOS runtime overhead}
\label{table:overhead}
\begin{tabular}{p{5.75cm} c}
\toprule
\textbf{Operation}                                           &       \textbf{Cost} \\ 
\midrule

JIT: storing TCB and other states to MRAM                                       &       1.11 ms \hl{(0.104 mJ)}\\ 
JIT: storing taskset's data to MRAM                                      &       1.46 ms \hl{(0.137 mJ)}\\ 
JIT: restoring states \& data from MRAM on reboot & 0.13 ms \hl{(0.013 mJ)}\\
\midrule
Max. JIT overhead during experiments & 0.088\% \\ 
Max. scheduler overhead during experiments & 0.001\% \\ 
Max. number of power cycles during experiments      &       105 \\ 
\midrule
MRAM vs SRAM: write operation power increase        & 53\%       \\ 
MRAM vs SRAM: read operation power increase        & 17\%       \\ 
\bottomrule

\end{tabular}%
\end{table}




\subsection{Analytical Experiments}
In this section, we compare the schedulability of our scheduler with Rtag~\cite{MohsenIOTJ2021}, which performed the best among other methods in the previous section. We exclude other related works from this comparison because they either lack schedulability analysis, rely on overly simplified real-time analysis equations (e.g., considering only taskset utilization), and reportedly perform worse than Rtag \cite{MohsenIOTJ2021}.

The experiments are conducted in MATLAB on a workstation equipped with an Intel 4.2GHz Core i7 CPU with 16GB of RAM. Since the SOTA method does not support the analysis of processing chains, we only generated single-task chains for comparison. Therefore, we refer to the generated chains as tasks in the rest of this section.

To determine the execution time of each task for our experiments, we choose the period randomly from 1s to 60s, find the task's utilization using the UUniFast method~\cite{UUnifast}, and then multiply this utilization by the task's period and round the result to the nearest positive value with one decimal point accuracy. In simpler terms, we determine the task execution time as $C_i = \max(\lfloor 10\cdot T_i\cdot U_i \rceil/10, 0.1)$. 

We begin by assessing the impact of discharging rates on schedulability. This experiment involves categorizing tasks into two distinct groups: those with high energy demands and those with low energy demands. For the high-energy demand tasks, the discharging rate is selected randomly from the range of 8 to 10, while for the low-energy demand tasks, it is chosen from 1 to 3. We introduce variations in the proportion of low-energy demand tasks, ranging from 0 to 100\%. For each configuration of the low task demand ratio, we create 1000 tasksets and calculate the average schedulability rate across these sets. In this experiment, the number of tasks in a taskset is set to 5 and the utilization of the taskset is randomly selected from a range of 0.1 to 0.9. The atomicity of each task (i.e., if the task is preemptible or non-preemptible) is also selected randomly. The priorities of tasks for each taskset are assigned based on the Rate-Monotonic method (i.e., tasks with lower periods get higher priority).
The charging rate is fixed to 3, and implicit deadlines are considered, i.e., $D_i=T_i$.

\begin{figure}[htb]
	\centering
	\includegraphics[width=0.4\textwidth]{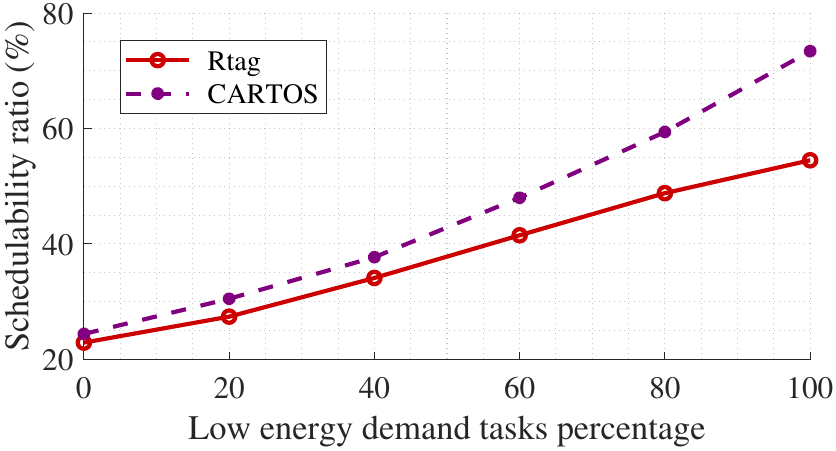}
	\caption{Scheduler performance with various discharging rates}
	\label{fig:simulation_comparison_mp}
\end{figure}

Fig.~\ref{fig:simulation_comparison_mp} shows that CARTOS outperforms SOTA by up to 20\% of schedulability.
This is mainly due to the unnecessary blocking time caused by lower-priority tasks in SOTA which treats all tasks non-preemptible even if some are fully computational tasks. The gap between the two methods increases with the number of low-energy demand tasks because the negative impact of the blocking time outweighs the energy constraint. 

Next, we evaluate the impact of utilization on taskset schedulability. In this experiment, the taskset utilization is chosen from 0.1 to 0.9 with 0.1 steps. For each taskset utilization, 1000 tasksets are generated and the average schedulability ratio of the total 1000 tasksets is reported. The number of tasks, period, and discharging rate of each task are chosen randomly from 3 to 8, from 1s to 60s, and from 1 to 10, respectively. We use the same method as the previous experiment to calculate the execution time for each of the tasks within a taskset.

\begin{figure}[htb]
	\centering
	\includegraphics[width=0.4\textwidth]{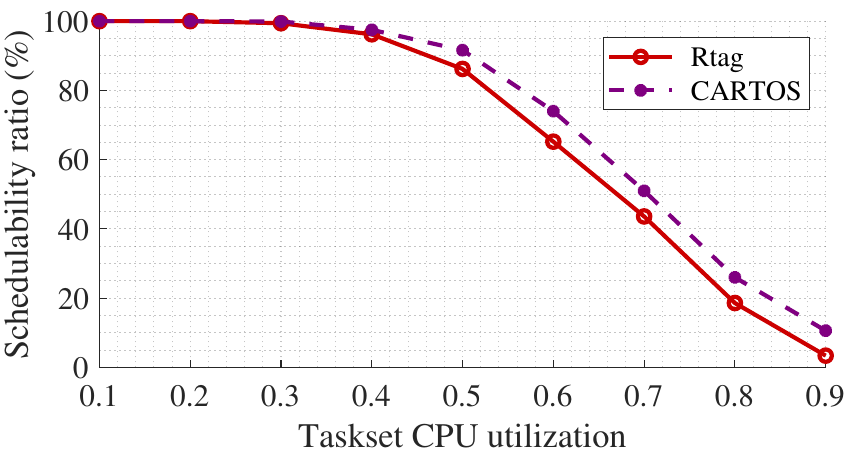}
	\caption{Scheduler performance with various utilization rates}
	\label{fig:simulation_comparison_u}
\end{figure}

Fig.~\ref{fig:simulation_comparison_u} shows the results. The schedulability improvement of CARTOS over SOTA tends to increase with the taskset utilization although the gap is smaller than in the previous experiment. From these results, we conclude that the energy constraint is indeed the major factor contributing to the schedulability performance in IPDs and the benefit of our mixed-preemption approach will be tremendous when the system has long-running computational tasks along with peripheral tasks.

\section{Conclusion}
\label{sec:conclusion}
In this paper, we introduced CARTOS, a charging-aware real-time operating system for intermittently-powered betteryless devices. CARTOS leverages a mixed-preemption scheduling model to facilitate correct and efficient execution of periodic sensing applications, comprised of a chain of computational and peripheral tasks. For computational tasks, it schedules non-preemptively while ensuring forward progress through just-in-time (JIT) checkpointing. For peripheral tasks, it guarantees atomicity by scheduling them non-preemptively and charging sufficient energy to complete their jobs before execution.
Furthermore, CARTOS includes energy management that computes the required charging time and utilizes a machine-learning-based energy predictor to update the harvesting rate to adapt to changing environmental conditions. 

We evaluated CARTOS through real system experiments as well as schedulability experiments. Our results showed that CARTOS outperforms previous methods by successfully scheduling periodic tasks regardless of the capacitor size and is able to schedule all higher priority jobs even in low charging situations with predictability. Furthermore, CARTOS  outperformed state-of-the-art by up to 20\% in schedulability across various experimental settings. 

For future work, we plan to expand the capabilities of CARTOS to accommodate a more diverse set of applications. This encompasses task synchronization and locking primitives, integrating with multi-core microcontrollers, and managing multiple energy buffers, all within the context of real-time operating systems and batteryless devices.







\bibliographystyle{IEEEtran}
\bibliography{Ref.bib}


\vskip -2\baselineskip plus -1fil

\begin{IEEEbiography}[{\includegraphics[width=1in,height=1.25in,clip,keepaspectratio]{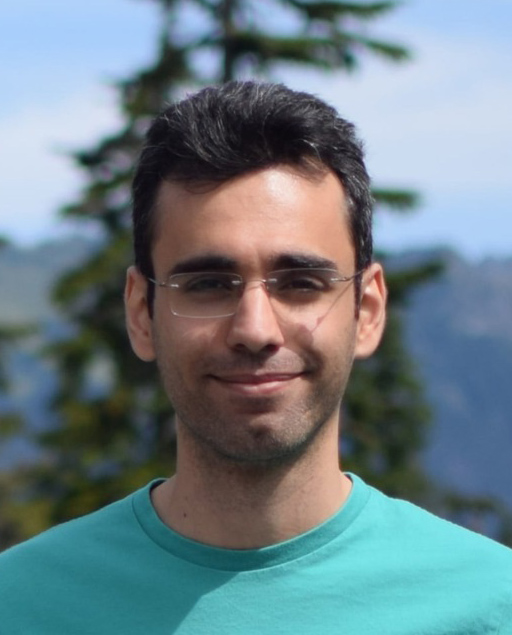}}]{Mohsen Karimi}
(Member, IEEE) received the Ph.D. degree in electrical and computer engineering from the University of California at Riverside, Riverside, CA, USA, in 2023. He is currently a Project Scientist with the Department of Electrical and Computer Engineering, University of California at Riverside, Riverside, CA, USA. His research interests include real-time task scheduling, embedded systems, cyber-physical systems, and machine learning, with a focus on designing energy harvesting devices, real-time task scheduling, resource management, and machine learning techniques for batteryless devices.
\end{IEEEbiography}

\vskip -2\baselineskip plus -1fil

\begin{IEEEbiography}[{\includegraphics[width=1in,height=1.25in,clip,keepaspectratio]{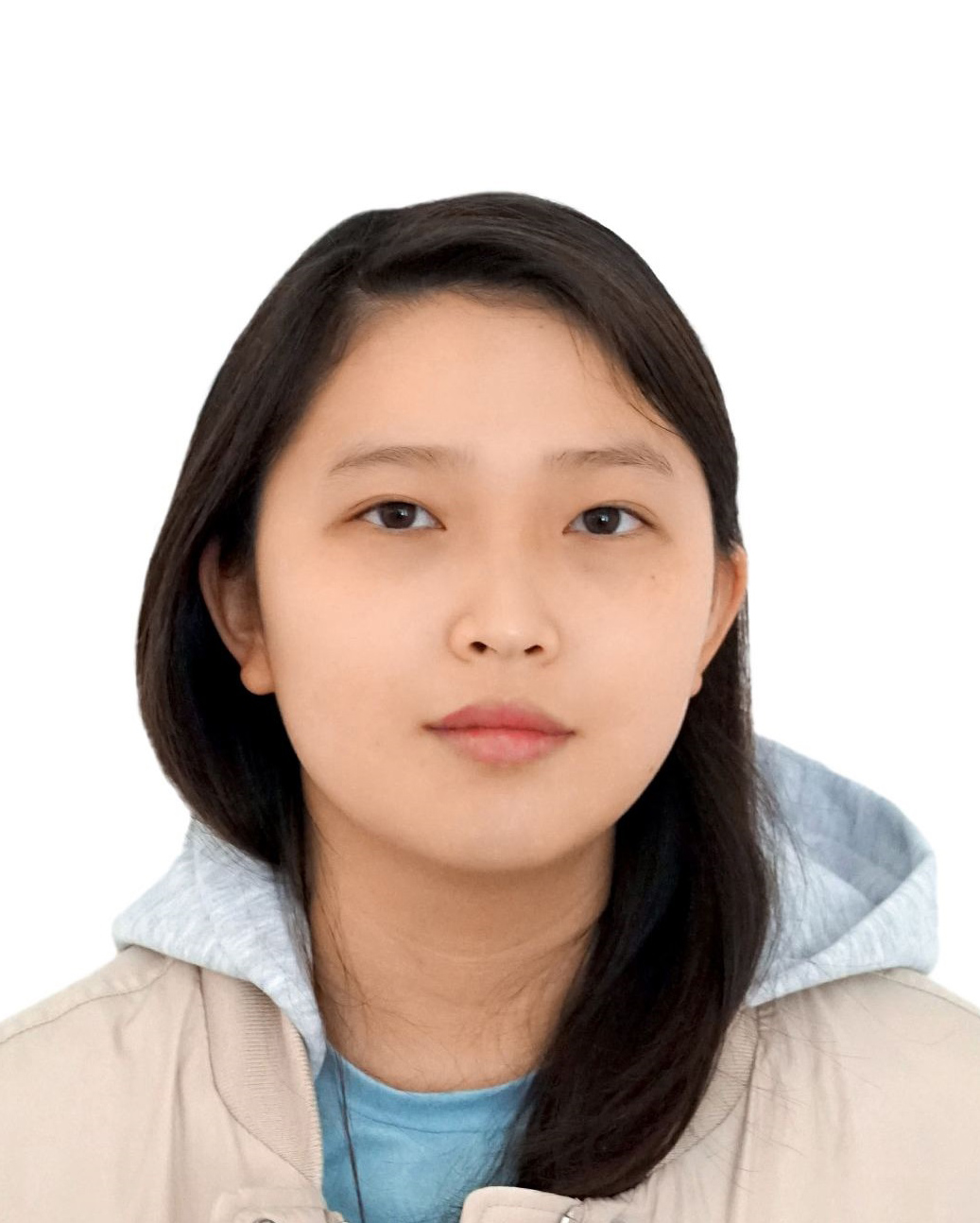}}]{Yidi Wang}
(Member, IEEE) 
Yidi Wang is currently an Assistant Professor in the Department of Computer Science and Engineering at Santa Clara University. Before joining SCU in 2024, she was a Postdoc in the Department of Electrical and Computer Engineering at the University of California, Riverside. She earned her Ph.D. in Electrical Engineering from UC Riverside in June 2023. Her primary research interests lie in real-time embedded and cyber-physical systems with a particular focus on real-time scheduling on heterogeneous systems in resource-constrained environments.
\end{IEEEbiography}

\vskip -2\baselineskip plus -1fil

\begin{IEEEbiography}[{\includegraphics[width=1in,height=1.25in,clip,keepaspectratio]{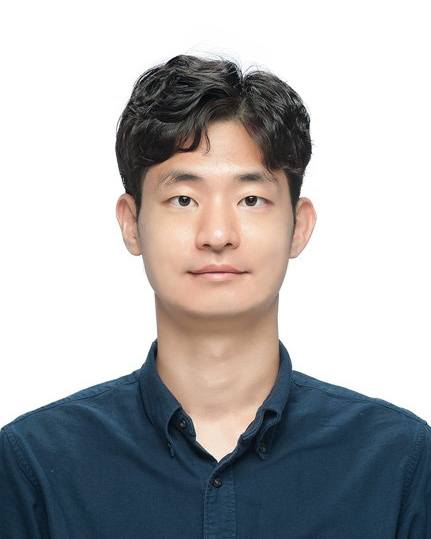}}]{Youngbin Kim}
(Member, IEEE) is a Senior Researcher in Electronics and Telecommunications Research Institute, Daejeon, South Korea. He received his B.S. degree in 2012 and Ph.D. degree in 2021, both from Yonsei University, Seoul, South Korea. His research is in the area of embedded systems, with a special interest in intermittent computing, neural network acceleration, and reliability.
\end{IEEEbiography}

\vskip -2\baselineskip plus -1fil

\begin{IEEEbiography}[{\includegraphics[width=1in,height=1.25in,clip,keepaspectratio]{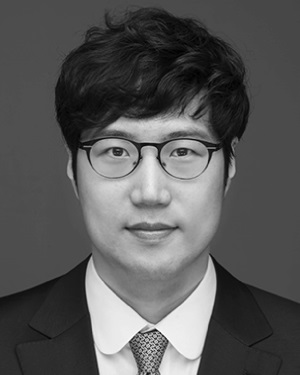}}]{Yoojin Lim}
(Member, IEEE) received the B.S. degree in computer science and computer engineering and the M.S. degree in information technology from Handong Global University, South Korea, in 2004, and the Ph.D. degree in business information technology from Kookmin University, South Korea, in 2016. From 2004 to 2008, he was a Senior Software Engineer with LG Electronics, South Korea. Since 2017, he has been a postdoctoral researcher and senior researcher with Electronics and Telecommunications Research Institute, Daejeon, South Korea, and he is currently a project leader of autonomous energy-driven computing system software research. His research interests include energy-driven intermittent computing, embedded software, hypervisors, cyber-physical systems, and mobility systems.
\end{IEEEbiography}

\vskip -2\baselineskip plus -1fil

\begin{IEEEbiography}[{\includegraphics[width=1in,height=1.25in,clip,keepaspectratio]{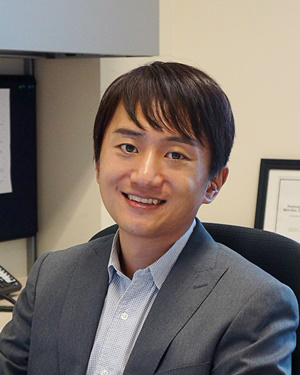}}]{Hyoseung Kim}
(Member, IEEE) received the Ph.D. degree in electrical and computer engineering from Carnegie Mellon University, Pittsburgh, PA, USA, in 2016. He is an Associate Professor with the Department of Electrical and Computer Engineering, University of California at Riverside, Riverside, CA, USA. His research interests are in real-time embedded and cyber-physical systems, especially at the intersection of systems software, parallel computing platforms, and analytical techniques. 
Dr. Kim is a recipient of the NSF CAREER Award, the Fulbright Scholarship Award, and the Best Paper Awards from RTAS and RTCSA. For more information, please visit \url{http://www.ece.ucr.edu/~hyoseung/}.
\end{IEEEbiography}

\end{document}